\documentclass{article}

\usepackage[margin=1in]{geometry}
\usepackage{url}
\usepackage[utf8]{inputenc}

\usepackage{graphicx}
\usepackage{amsmath}
\usepackage[labelformat=simple]{subfig}

\newcommand{\Hil}{{\mathcal H}}
\newcommand{\Ell}{{\mathcal L}}

\newcommand{\Boltz}{ k_{\rm\scriptscriptstyle B}}

\newcommand{\Tr}{{\rm Tr}}
\newcommand{\Ran}{{\rm Ran}}
\newcommand{\Ker}{{\rm Ker}}

\newcommand{\J}{{J\vphantom{\overline{J}}}}
\newcommand{\Jbar}{{\overline{J}}}
\newcommand{\ddt}[1]{{\frac{\displaystyle {\rm d}#1}{\displaystyle {\rm d}t}}}

\newcommand{\sq}{{\sqrt{\rho}}}
\newcommand{\sqnd}{{\sqrt{\rho_{\rm nd}}}}

\newcommand{\cov}[2]{{\langle\Delta #1\Delta #2\rangle}}
\newcommand{\sqcov}[2]{{\sqrt{\sigma_{\!#1\!#2}}}}
\newcommand{\scov}[2]{{\sigma_{\!#1\!#2}}}
\newcommand{\sccov}[2]{{\sigma^2_{\!#1\!#2}}}
\newcommand{\ceta}[2]{{\eta_{#1\!#2}}}
\newcommand{\cseta}[2]{{\eta^*_{#1\!#2}}}
\newcommand{\cceta}[2]{{\eta^2_{#1\!#2}}}
\newcommand{\com}[2]{{\langle [ #1, #2]/2i\rangle}}
\newcommand{\mean}[1]{{\langle #1\rangle}}

\newcommand{\citen}[1]{{\cite{#1}}}

\newcommand{\N}{{\mathbf N}}
\newcommand{\mmu}{{\boldsymbol{\mu}}}
\newcommand{\nnu}{{\boldsymbol{\nu}}}

\newcommand{\Pen}{{P_{\!e_{\!n}}}}
\newcommand{\Pem}{{P_{\!e_{\!m}}}}

\newcommand{\Petwo}{{P_{\!e_{2}}}}
\newcommand{\Pethree}{{P_{\!e_{3}}}}

\newcommand{\rFM}{{r_{\!F\!M}}}
\newcommand{\cFH}{{c_{F\!H}}}
\newcommand{\cMH}{{c_{\!M\!H}}}
\newcommand{\cPenH}{{c_{\Pen\!\!H}}}
\newcommand{\rPenM}{{r_{\Pen\!\!M}}}
\newcommand{\rAM}{{r_{\!A\!M}}}
\newcommand{\rSM}{{r_{\!S\!M}}}

\newcommand{\rSHprime}{{r_{\!S\!H'}}}
\newcommand{\rrFH}{{r^2_{\!F\!H}}}

\newcommand{\ccFH}{{c^2_{\!F\!H}}}
\newcommand{\rrSM}{{r^2_{\!S\!M}}}

\newcommand{\rrSHprime}{{r^2_{\!S\!H'}}}
\newcommand{\rrAM}{{r^2_{\!A\!M}}}

\newcommand{\be}{\begin{equation}}
\newcommand{\ee}{\end{equation}}
\newcommand{\bea}{\begin{eqnarray}}
\newcommand{\eea}{\end{eqnarray}}
\newcommand{\fr}{\frac}

\title{Time--Energy and Time--Entropy Uncertainty Relations in Nonequilibrium Quantum Thermodynamics under Steepest-Entropy-Ascent Nonlinear Master~Equations}

\author{Gian Paolo Beretta\\
	\small Università di Brescia, Italy
}

\date{published in Entropy, Vol. 21, 679 (2019)\\
	\url{http://dx.doi.org/10.3390/e21070679}}

\begin{document}
	\maketitle

\abstract{In the domain of nondissipative unitary Hamiltonian dynamics, the well-known Mandelstam--Tamm--Messiah time--energy uncertainty relation
	$\tau_{F}\Delta_H\ge \hbar/2$ provides a general lower
	bound to the characteristic time $\tau_F =\Delta_F/|{\rm d} \mean{F}/dt|$ with which the mean value of a generic quantum observable $F$ can change with respect to the width $\Delta_F$ of its uncertainty distribution (square root of  $F$ fluctuations). A useful practical consequence is that  in unitary dynamics the states with longer lifetimes are those with smaller energy uncertainty $\Delta_H$ (square root of  energy fluctuations). Here we show that when unitary evolution is complemented with a steepest-entropy-ascent model of dissipation, the resulting nonlinear master equation entails that these lower bounds get modified and depend also on the entropy uncertainty  $\Delta_S$ (square root of  entropy fluctuations). For example, we obtain the time--energy-and--time--entropy uncertainty relation $(2\tau_{F}\Delta_H/ \hbar)^2+
	(\tau_{F}\Delta_S/\Boltz\tau)^2 \ge 1$ where $\tau$ is a
	characteristic dissipation time functional that for each given state defines the strength of the nonunitary,	steepest-entropy-ascent part of the assumed master equation. For purely dissipative dynamics this reduces to the time--entropy uncertainty relation $\tau_{F}\Delta_S\ge \Boltz\tau$, meaning that  the nonequilibrium dissipative states with longer lifetime are those with smaller entropy uncertainty $\Delta_S$.
}

\section{Introduction}

Recent advances in quantum information and quantum thermodynamics (QT) have increased the importance of estimating the lifetime of a given quantum state, for example to engineer decoherence correction protocols aimed at entanglement preservation. In the same spirit as fluctuation theorems that allow to estimate some statistical features of the dynamics from suitable state properties, also the Mandelstam--Tamm--Messiah  time--energy uncertainty relations (MTM-TEURs) have been long known to provide bounds on lifetimes of quantum decaying states under Hamiltonian (non-dissipative) evolution.  For practical applications, however,  such bounds are insufficient when Hamiltonian dynamics must be complemented by models of dissipation and decoherence. 

The time--energy uncertainty relation  has remained an open and at
times controversial issue throughout the history of quantum
theory. Several reviews are available on the pioneering
discussions  and the subsequent developments \cite{Aharonov61,Allcock69_1,Allcock69_2,Allcock69_3,Bauer78,Dodonov80,Busch90_1,Busch90_2,Landauer91,Pfeifer95,Hu95,Hilgevoord96,Hilgevoord98,Trifonov02,Dodonov02}. In the present paper, we are motivated 
by the past two decades of important advancements in our understanding of the general structure of dynamical models for non-equilibrium thermodynamics, including non-equilibrium quantum thermodynamic models. Such revival has been prompted and
paralleled by a steady advancement of experimental
techniques dealing with single ion traps \cite{Sackett00,Kielpinski02}, qubits
\cite{Ekert02,Bovino05}, neutron interferometry \cite{Shull80,Martinez04}, and a countless and
growing number of other quantum-information developments since then, e.g., nonlinear quantum metrology \cite{DemkowiczDobrzanski17,Beau17}. 
Within these applications, TEURs can provide useful information and practical bounds for parameter estimation. But since dissipation and decoherence are often the limiting factors, there is a need to generalize the  MTM-TEURs to frameworks where microscopic few-particle quantum setups exhibit non-unitary dissipative dynamical behavior.

A recent review paper on the physical significance of TEURs provides 300 references and the following conclusion  \cite{Dodonov15}: ``We have shown that the area of energy--time uncertainty relations continues to attract
	attention of many researchers until now, and it remains alive almost 90 years after
	its birth. It received a {new breath} in the past quarter of century due to the actual
	problems of quantum information theory and impressive progress of the experimental
	technique in quantum optics and atomic physics. It is impossible to describe various
	applications of the TEURs to numerous different physical phenomena in this minireview.''

The main objective in the present paper is to extend  the time--energy uncertainty relations to the framework of dissipative quantum dynamical systems. But differently from the most popular and traditional model of dissipation in open quantum systems, which is based on the well-known   Kossakowski--Lindblad--Gorini--Sudarshan (KLGS) master equations \cite{Kraus71,Kossakowski72_1,Kossakowski72_2,Ingarden75,Lindblad76,Gorini76,Spohn76,Spohn78,Asorey09}, we assume the less known locally steepest-entropy-ascent (LSEA) model of dissipation. We make this choice not only to avoid some drawbacks  (outlined in more details in Appendix \ref{AppendixA})  of the KLGS master equation from the point of view of full and strong consistency with the general principles of thermodynamics, causality, and far non-equilibrium, but more importantly because we have shown in References \cite{PRE14,PRE15} that the LSEA principle---by  providing the minimal but essential  elements of thermodynamic consistency, near as well as far from stable (maximal entropy) equilibrium states---has the potential to unify all the successful frameworks of non-equilibrium modeling, from kinetic theory to chemical kinetics, from stochastic to mesoscopic to extended irreversible thermodynamics, as well as the metriplectic structure or, in more recent terms, the General Equation for Non-Equilibrium Reversible-Irreversible Coupling (GENERIC) structure.  In addition, it is noteworthy that  a particular but broad class of KLGS master equations has been recently shown to fall into a LSEA (entropic gradient flow) structure  \cite{Mittnenzweig17,Kantner17}, and hence some of the TEURs we derive here hold also for such class of models. 

Steepest-entropy-ascent (SEA) nonlinear master equations  have proved to be effective tools to model dissipative dynamics, thermalization, transport processes and, in general, entropy production in a wide range of frameworks of non-equilibrium thermodynamics. In essence, SEA models are  explicit implementations of the general principle of maximum local entropy production. In recent mathematical terms, SEA models are entropic gradient flows. From the fundamental point of view, the general structure and nonlinearity of the  SEA master equations are instrumental to providing strong compatibility with the second law of thermodynamics by guaranteeing, within the model, the Hatsopoulos--Keenan statement of existence and uniqueness of the stable equilibrium (maximum entropy) states. Here we focus on the quantum thermodynamic modeling framework of application and show how the entropy production modifies the usual TEURs. 

The usual
time--energy uncertainty relation---as interpreted according to the
Mandelstam--Tamm--
Messiah intrinsic-time approach \cite{Messiah76,Mandelstam45}
based on unitary Hamiltonian dynamics---is modified by the presence
of a maximally dissipative term in the dynamical law, which models
at the single- or few-particle quantum level the so-called maximum entropy
production principle (MEPP) \cite{Ziegler58,Swenson89,Struchtrup98, Dewar05,Martyushev06,Martyushev10,Martyushev13,Martyushev14}. TEURs
obtained in other frameworks \cite{Bialynicki75,Gislason85,Kobe94,Majernik97,Aharonov00,Aharanov01,Busch01,Brunetti02,Gillies05} such as attempts to
define time or ``tempus'' operators, entropic uncertainties, and
measurement times are beyond our scope here.

The class of MEPP  master equations we designed in References \cite{thesis,Cimento1,Cimento2,Nature,Maryland86,ROMP} is suitable to   model dissipation phenomenologically not only in  open quantum
systems in contact with macroscopic baths, but also in closed isolated systems, as well as strongly coupled and entangled composite systems (references below).
These master equations are capable to
describe the natural tendency of any
initial nonequilibrium state (read: density operator) to relax
towards canonical or partially-canonical thermodynamic
equilibrium (Gibbs state), i.e., capable of describing the irreversible tendency
to evolve towards the highest entropy state compatible with the
instantaneous mean values of the energy (and possibly other
constants of the motion and other constraints). They do so by preserving exactly the conserved properties while pulling  every nonequilibrium state in the SEA direction with respect to the local dissipation metric that is part of the nonequilibrium description of the system \cite{PRE14}. This dissipative tendency is simultaneous and in competition with the usual non-dissipative
Hamiltonian unitary evolution.

Our original approach---when understood as an attempt to develop thermodynamically consistent modeling approaches that merge mechanics and thermodynamics following Hatsopoulos and Gyftopoulos \cite{HG3,HG1,HG2,HG4}---can perhaps be considered a first pioneering ``resource theory'' of quantum thermodynamics equipped with a nonlinear dissipative dynamical structure capable to describe relaxation even from arbitrarily far from equilibrium and to entail the second law as a theorem of the dynamical law. Several other pioneering aspects of  QT resource theories were present in References \cite{HG3,HG1,HG2,HG4}. For example, the energy versus entropy diagram to represent nonequilibrium states in the QT framework, first introduced in Reference \cite{HG3}, has recently found interesting applications  in \cite{Oppenheim13}. Again, it provides definitions and expressions for adiabatic availability and available energy with respect to a heat bath, work element, heat interaction, etc. which are currently discussed intensely in the QT community. It must also be mentioned that this first QT resource theory  was proposed in years when talking of quantum thermodynamics was considered heresy by the orthodox physical community. Considering that it is little cited and still not well known, we give more details below and in Appendix \ref{AppendixB}.  

We provide in the two appendices a brief review of some practical and conceptual issues of the prevailing model of irreversibility, and a  discussion
of the original motivation that lead us to develop a quantum
maximal entropy production formalism. We do not repeat here the geometrical derivations of
our nonlinear MEPP dynamical law, nor the discussions of its many
intriguing mathematical--physics implications, because they are available in many previous papers. Here, we simply
adopt that master equation without derivation, and focus on its
consequences related to TEURs, illustrated also
by some numerical simulations.  Quantum statistical mechanics and quantum thermodynamics practitioners have so far essentially dismissed and ignored our class of SEA master equations on the basis that they do not belong to the standard class of KLGS master equations and hence cannot be the correct description of the reduced dynamics of a system in interaction with one or more thermal baths. However, at least when used as phenomenological modeling tools,  SEA master equations have recently proved \cite{Cano15,Smith16,Tabakin17,Yamada18,Li18,Kusaba17,Militello18,Yamada19,Yamada19PRE} to offer in a variety of fields  important advantages of broader or complementary applicability for the description and correlation of near- and far-non-equilibrium behavior.

In the quantum  framework, the local state of a subsystem is represented by the local density operator $\rho$ and its lifetime may be characterized by the intrinsic characteristic times $\tau_{F}$ of the dynamical
variables associated with the linear functionals $\Tr(\rho F)$. If the local dynamics is non-dissipative and described by the usual unitary evolution, we  show below that the Heisenberg--Robertson inequality entails the usual
MTM-TEURs
$\tau_{F}\Delta_H\ge \hbar/2$, while the Schroedinger inequality entails sharper and more general exact
TEURs [Equation\ (\ref{nondissTE})].

For simultaneous
unitary$+$dissipative dynamics, the usual
TEUR is expectedly 
replaced
by less restrictive relations and additional characteristic times acquire physical significance. In particular,  we focus our attention to the  characteristic time associated with the rate of change of the  von Neumann  entropy functional
$-\Boltz\Tr(\rho\ln\rho)$.  For unitary$+$LSEA evolution, in Section \ref{Example} [Equation\ (\ref{genunc6})] we obtain an interesting time--energy
and time--entropy uncertainty relation $(2\tau_{F}\Delta_H/ \hbar)^2+
(\tau_{F}\Delta_S/\Boltz\tau)^2 \ge 1$ where $\tau$ is the main dissipation time  that defines the strength of the
dissipative component of the assumed dynamical law. 
With the help of numerical simulations,
we illustrate this relation
and several other even more precise uncertainty relations, that in the framework of QT resource theories may have a useful application in quantifying the lifetime of quantum states. 

The structure of the paper is outlined at the end of the next section, where we first introduce the particular class of  nonlinear dissipative quantum master equations on which we restrict our attention in the first part of the paper.

\section{\label{Structure}Assumed Structure of the Nonlinear Dissipative Quantum Master Equation}

Let $\Hil$ (${\rm dim}\Hil \le \infty$) be the Hilbert space and
$H$ the Hamiltonian operator that in standard Quantum Mechanics we
associate  with a given isolated (or adiabatic, 
see below)
and uncorrelated system. We assume that the quantum states are
one-to-one with the linear hermitian operators $\rho$ on $\Hil$
with $\Tr(\rho)=1$ and $\rho\ge \rho^2$, and we assume a dynamical
equation of the form 
\begin{equation}
\label{rhodot}\ddt{\rho}=\rho\,E(\rho)+ E^\dagger(\rho)\,\rho \ ,
\end{equation}
where $E(\rho)$ is an operator-valued function of $\rho$ that we
may call  the ``evolution generator'' which may in general be
non-hermitian and nonlinear in $\rho$, but must be such as to
preserve $\rho$ unit trace and non-negative definite. Without loss
of generality, we write $E=E_+ +iE_-$ where $E_+=(E+E^\dagger)/2$
and $E_-=(E-E^\dagger)/2i$ are hermitian operators. Then, the
dynamical law takes the form
\begin{equation}
\ddt{\rho}=-i\,[E_-,\rho]+ \{E_+,\rho\} \ ,
\end{equation}
where $[\,\cdot\,,\,\cdot\,]$ and  $\{\,\cdot\,,\,\cdot\,\}$ are
the usual commutator and anticommutator, respectively. In Appendix \ref{AppendixA} 
we discuss the reasons why we adopt this form, and exclude
terms like $V(\rho)\,\rho\,V(\rho)$ which appear instead in the
the celebrated KSGL class of (linear)
quantum master equations.

In preparation for our SEA  construction in Section \ref{Example}, we assume
$E_-=H/\hbar$ (independent of $\rho$), where $H$ is the Hamiltonian operator of the system and $\hbar$  the
reduced Planck constant, and rewrite $E_+$ as $E_+=\Delta
M(\rho)/2\Boltz\tau$ where $\Boltz$ is the Boltzmann
constant, $\tau$  a positive constant (or state functional) that in the SEA framework we will interpret as an intrinsic dissipation time of the system, because it essentially fixes the rate at which the state evolves along the path of SEA in state space,
 and $\Delta M(\rho)$  a
hermitian operator-valued nonlinear function of $\rho$ that we
call the ``nonequilibrium Massieu operator'' and until Section \ref{Example} we do not define explicitly, except for the assumption that it satisfies the condition  \begin{equation}\label{preservetrace1}
\Tr[\rho\Delta M(\rho)]=0
\end{equation} as well as the condition that it preserves the nonnegativity of $\rho$ (both forward and backwards in time!). As a result, Equation\ (\ref{rhodot}) takes the form
\begin{equation}
\label{rhodotHM}\ddt{\rho}=-\frac{i}{\hbar}[H,\rho]+
\frac{1}{2\Boltz\tau }\{\Delta M(\rho),\rho\} \, .
\end{equation}

Let us note that
as in standard unitary dynamics, we say  that the particle is
either isolated or adiabatic, respectively, if the Hamiltonian operator $H$ is
either time independent or time dependent, for example, through one or more external control parameters.

In Section \ref{Example}, we will consider for $\Delta M(\rho)$ the  explicit SEA form for the simplest case, first proposed in \cite{thesis,Cimento1,Maryland86}. Reference \cite{Cimento2} proposed also a general LSEA form for a composite quantum system, which will not be considered here, but has clear applications in the description of decoherence and lifetime of entanglement (see Reference \cite{Cano15}).

In the present paper we limit the discussion to the derivation of general inequalities, and to
illustrative considerations and a numerical example valid within the
simplest framework of steepest-entropy-ascent conservative dynamics.
The application to structured composite systems based on our LSEA version \cite{Cimento2,Cano15} of operator $\Delta M(\rho)$ will be discussed elsewhere.

The specific physical interpretations of the uncertainty relations
that follow from  dynamical law (\ref{rhodotHM}) will depend on
the theoretical or modeling context in which such time evolution is
assumed. For example, the problem of designing well-behaved nonlinear
extensions of the standard unitary  dynamical law of quantum
mechanics has been faced in the past few decades with a variety of
motivations, and is recently seeing a vigorous revival in connection
with questions about the foundations of quantum mechanics and the need for thermodynamically sound phenomenological models (recently referred to as ``resource theories'' \cite{Oppenheim13})  that arise
from the current developments of quantum information technologies and
related single-particle and single-photon experiments to test quantum computing components and devices and fundamental questions  about entanglement, decoherence, nonlocality,
and measurement theory.

In our original development \cite{thesis,Cimento1,Cimento2,Nature,Maryland86}, Equation (\ref{rhodotHM})
was designed as part of an ad-hoc fundamental dynamical postulate
needed \cite{SantaFe84_1,Paris87,Taormina92,Leiden06,MPLA1,MPLA2} to complete the Hatsopoulos-Gyftopoulos attempt \cite{HG3,HG1,HG2,HG4,Park78,Simmons81}
to unify mechanics and thermodynamics into a  generalized quantum theory by building the Hatsopoulos--Keenan statement of the
second law \cite{HK,Book} directly into the microscopic level of
description. In particular, the key ansatz  in References \cite{HG3,HG1,HG2,HG4}  is the
assumption that even for a strictly isolated system, there exists a
broad class of genuine states (homogeneous preparations, in von
Neumann language \cite{thesis,MPLA2,vonNeumann,ParkState}) that require non-idempotent
density operators, i.e., such that $\rho^2\ne\rho$. Two decades later, this
ansatz has been re-proposed in Reference \cite{Bona}, and our
nonlinear dynamical Equation~(\ref{rhodotHM})  has been re-discovered
and studied in References \cite{Gheorghiu1,Gheorghiu2,Gheorghiu3,Gheorghiu4}, where it is shown to be
well-behaved from various perspectives including a relativistic
point of view. An important feature is that it entails non-unitary
evolution only for non-idempotent ($\rho^2\ne\rho$) density
operators, whereas for idempotent ($\rho^2 =\rho$) density operators
it entails the standard unitary evolution (see, e.g., References~\citen{Cimento1,PRE06}).

However, the present results are valid also in any other framework, theoretical discussion, modeling context, or resource theory   whereby---for example  to study decoherence,
dissipation, quantum thermal engines, quantum refrigerators, and so on---the usual Liouville-von
Neumann equation for the density operator is modified, linearly or nonlinearly, into form (\ref{rhodotHM}). 

Since many of the relations we derive here are valid and
nontrivial in all these contexts, in \mbox{Sections
\ref{General}--\ref{Occupation}} we begin by presenting the  results that do
not depend on assuming a particular form
of operator $\Delta M(\rho)$. Thus, independently of the interpretation, the context of application, and the specific form of master Equation (\ref{rhodotHM}), the uncertainty relations derived in
the first part of the paper  extend the usual relations to the far non-equilibrium
domain and in general to all non-zero-entropy states. 

 In Sections
\ref{Example} and \ref{Numerical}, to fix ideas and be able to present numerical results and qualitative 
considerations, we specialize the analysis to  the simplest nontrivial form of Equation (\ref{rhodotHM}) that implements our
conservative steepest-entropy-ascent  dynamical ansatz, namely, a model
for irreversible relaxation of a four-level qudit.

Appendix \ref{AppendixA}  discusses our reasons for not considering, in the
present context, the extension of our results to a full
Kossakowski-Lindblad form of the evolution equation.

Appendix \ref{AppendixB}  gives a brief review of the original motivations that lead us to develop the SEA and LSEA formalism in the early quantum thermodynamics scenario, and of the subsequent developments that in recent years have shown how the locally steepest-entropy-ascent principle not only gives a clear, explicit, and unambiguous meaning to the MEPP but it also constitutes  the heart of (and essentially unifies) all successful theories of nonequilibrium.   

\section{\label{General}General Uncertainty Relations}
We consider the space $\Ell (\Hil)$ of linear operators on $\Hil$
equipped with  the real scalar product
\begin{equation}
\label{Scalar} (F|G) = \Tr (F^\dagger G + G^\dagger F)/2 =(G|F) \
,
\end{equation}
and the real antisymmetric bilinear form
\begin{equation}
\label{Commutator} (F\backslash G) = i\,\Tr (F^\dagger G -
G^\dagger F)/2 =-(G\backslash F)= (F|iG)\ ,
\end{equation}
so that for any  hermitian $F$ in $\Ell (\Hil)$
the corresponding mean-value state functional can be written as
$\mean{F}= \Tr (\rho F)=\Tr (\sq F\sq)=(\sq|\sq F)$, and can
therefore be viewed as a functional of $\sq$, the square-root
density operator, obtained from the spectral expansion of $\rho$
by substituting its eigenvalues with their positive square roots.
When $\rho$ evolves according to Equation (\ref{rhodot}) and $F$ is time-independent, the rate of
change of $\Tr (\rho F)$ can be written as
\begin{equation}
 \label{rateF}  {\rm d} \Tr (\rho F)/{\rm d}t = \Tr (
F\,{\rm d} \rho /{\rm d}t)=2\left.\left(\sq F\right|\sq
E(\rho)\right) \ .
\end{equation}

In particular, for the evolution Equation (\ref{rhodot}) to be well
defined, the functional $\Tr (\rho I)$ where $I$ is the identity on
$\Hil$ must remain equal to unity at all times; therefore, $ {\rm d}
\Tr (\rho I)/{\rm d}t =2\left.\left(\sq I\right|\sq E(\rho)\right)=0
$ or, equivalently,  in view of Equation\ (\ref{rhodotHM}), Equation\ (\ref{preservetrace1}) rewrites as
\begin{equation}
 \label{rateM}\left.\left(\sq \right|\sq \Delta M(\rho)\right) =0\ .
\end{equation}

For $F$ and $G$ hermitian in $\Ell (\Hil)$,  we introduce the
following shorthand notation
\begin{eqnarray}
\label{Delta}\Delta F&=&F- \Tr(\rho F)I \ ,
\end{eqnarray}
\vspace{-12pt}
\begin{equation}
\begin{array}{lll}
\label{Cov}\scov{F}{G}&=&\cov{F}{G}= (\sq \Delta F|\sq \Delta G)
\\ 
&=&\tfrac{1}{2} \Tr(\rho\{\Delta F,\Delta
G\})=\scov{G}{F}\ ,\\
\end{array}
\end{equation}
\vspace{-12pt}
\begin{eqnarray}
 \Delta_F&=&\sqcov{F}{F}=\sqrt{\cov{F}{F}}\
,
\end{eqnarray}
\vspace{-12pt}
\begin{equation}
\begin{array}{lll}
  \label{Covi}\ceta{F}{G}&=&\com{F}{G}= (\sq
\Delta F\backslash \sq \Delta G) \\
&=& \frac{1}{2i}
\Tr(\rho[ F, G])=\cseta{F}{G}=-\ceta{G}{F}\ ,
\end{array}
\end{equation}

 For
example, we may write the rate of change of the mean value of a
time-independent observable $F$ as
\begin{equation}
 \label{rateF2} \frac{{\rm d}\Tr (\rho F)}{{\rm d}t} =
 \frac{\com{F}{H}}{\hbar/2}
 + \frac{\cov{F}{M}}{\Boltz\tau}=\frac{\ceta{F}{H}}{\hbar/2}
 + \frac{\scov{F}{M}}{\Boltz\tau} \ ,
\end{equation}
from which we see that not all operators $F$ that commute with $H$
correspond to constants of the motion, but only those for which
$\cov{F}{M}=0$, i.e., such that $\sq\Delta F$ is orthogonal to
both $i\sq\Delta H$ and $\sq\Delta M$, in the sense of scalar
product (\ref{Scalar}). For an isolated system, conservation of
the mean energy functional $\Tr(\rho H)$ requires an operator
function $\Delta M(\rho)$ that maintains $\sq\Delta M$ always
orthogonal to $\sq\Delta H$, so that $\cov{H}{M}=0$  for every
$\rho$.

From the Schwarz inequality, we readily verify 
 the following generalized
Schr\"odinger uncertainty relation
\begin{equation}
 \label{Sinequality} \cov{F}{F}\cov{G}{G}\ge \cov{F}{G}^2+  \com{F}{G}^2\
 ,
\end{equation}
usually written in the form
$\sqrt{\scov{F}{F}\scov{G}{G}-\sccov{F}{G}}\ge |\ceta{F}{G}|$. It follows
from the Cauchy--Schwarz inequality $( f,f)( g,g ) \ge
|( f,g ) |^2$ and the identity $|( f,g ) |^2=(f|g)^2+(f\backslash
g)^2$ where $ (f|g)=[( f,g )+( g,f ) ]/2$, $(f\backslash g)=i[(
f,g )-( g,f )]/2$, and $f$, $g$ are vectors in some complex
Hilbert space (strict equality iff $f=\lambda\, g$ for some scalar
$\lambda$). In the space $\Ell'(\Hil)$ of linear operators on $\Hil$
equipped with the complex scalar product $(f,g)=\Tr(f^\dagger g)$,
we note that $(f,f)=(f|f)$
and obtain the inequality
$(f|f)(g|g)\ge (f|g)^2+(f\backslash g)^2$ and hence inequality
(\ref{Sinequality}) by setting $f=\sq\Delta F$ and $g=\sq\Delta
G$. Note that the strict equality in (\ref{Sinequality}) holds iff
$\sq\Delta F=\lambda\sq\Delta G$ for some scalar $\lambda$ (in which case we
have $\com{F}{G}=0$ iff either $\lambda^*=\lambda$ or $\sq\Delta F=0$ or
both). This proof was given as footnote 7 of Reference~\citen{ArXiv05}. For Schroedinger's original proof and an alternative one see Reference~\citen{Rigolin15}.  Relation (\ref{Sinequality}) is
a generalization of the inequality first appeared in~\cite{Schroedinger30,Robertson30} and later generalized in~\cite{Robertson34} to the form  $\det{\sigma}=\scov{F}{F}\scov{G}{G}
-\sccov{F}{G} \ge \det{\eta}=\cceta{F}{G}=\cceta{G}{F}$, suitable
for generalizations to more than two observables. Early
proofs of relation (\ref{Sinequality}) were restricted to pure
state operators ($\rho^2=\rho$). To our knowledge, the earliest
proof valid
for general (mixed and pure) states $\rho$  is that in  \cite{Dodonov80}. 	
For further inequalities in the case of position and
momentum operators see \cite{Dodonov02} and references therein. 	Notice also that by using our proof of the Schrodinger
inequality (\ref{Sinequality}), just given above, Relation (22) of the main theorem in
the review paper \cite{Pfeifer95} can be made sharper and read $
|\Tr(R[A,B])|^2+|\Tr(R\{A,B\})-2\sum_n\lambda_n\Tr(P_nAP_nB)|^2 \le
4f^2(R,A)f^2(R,B) $.

Relation (\ref{Sinequality}) obviously entails the less precise
and less symmetric Heisenberg-Robertson uncertainty relation
\begin{equation}
 \label{Rinequality} \cov{F}{F}\cov{G}{G}\ge  \com{F}{G}^2\ ,
\end{equation}
usually written in the form $\Delta_F\Delta_G\ge |\ceta{F}{G}|$.

For further compactness, we introduce the notation
\begin{eqnarray}
\label{rFG} r_{FG}&=&\scov{F}{G}\big/\sqrt{\scov{F}{F}\scov{G}{G}}
\ ,\nonumber\\ \label{cFG}
c_{FG}&=&\ceta{F}{G}\big/\sqrt{\scov{F}{F}\scov{G}{G}} \
,\end{eqnarray} where clearly, $r_{FG}$ represents the cosine of
the angle between the `vectors' $\sq \Delta F$ and $\sq \Delta G$
in $\Ell(\Hil)$, and $ r^2_{FG}\le 1$. Inequality
(\ref{Sinequality}) may thus be rewritten as
 \begin{equation}
 r^2_{FG}+c^2_{FG}\le 1 \end{equation} and
clearly implies \begin{equation}c^2_{FG}\le
\frac{1}{1+(r^2_{FG}/c^2_{FG})}\le 1-r^2_{FG}\le 1\
.\end{equation}

Next, for any hermitian $F$ we define the
 characteristic time of change  of the corresponding property defined by the  mean
value of the linear functional $\mean{F}= \Tr(\rho F)$  as follows
\begin{equation}
\tau_{F}(\rho)=\Delta_F\big/|{\rm d}\mean{F}/{\rm d}t| \
.\label{tauF}
\end{equation}

As is well known \cite{Aharonov61,Allcock69_1,Allcock69_2,Allcock69_3,Bauer78,Gislason85,Busch90_1,Busch90_2,Pfeifer95,Hilgevoord96,Hilgevoord98,Dodonov02,Messiah76,Mandelstam45}, $\tau_{F}$ represents
the time required for the statistical distribution of measurements
of observable $F$ to be appreciably modified, i.e., for the mean
value $\mean{F}$ to change by an amount equal to the width
$\Delta_F$ of the distribution.

Now, defining the nonnegative, dimensionless functional
\begin{equation}\label{ataudef}
a_\tau=\hbar\Delta_M\big/2\Boltz\tau\Delta_H \ ,
 \end{equation}
 we  rewrite (\ref{rateF2}) in the form
\begin{equation}
 \label{rateF3} {\rm d}\mean{F}/{\rm d}t =
 2\Delta_F\Delta_H\,(\cFH+a_\tau\,\rFM)/\hbar
 \end{equation}
and, substituting  into (\ref{tauF}), we obtain the general exact
uncertainty relation
\begin{equation}\label{exactTE}
\frac{\hbar/2}{\tau_{F}\Delta_H}= |\cFH+a_\tau\,\rFM| \ .
 \end{equation}
 
For~non-dissipative dynamics [$\Delta M(\rho)/\tau=0$], $a_\tau=0$, Equation\ (\ref{exactTE})
yields the time--energy uncertainty relations
 \begin{equation}\label{nondissTE} \frac{\hbar^2/4}{\tau^2_{F}\scov{H}{H}}
 =\ccFH\le
\frac{1}{1+(\rrFH/\ccFH)}\le 1-\rrFH\le 1\ ,\end{equation}
 which entail but are more
precise than the usual time--energy uncertainty relation, in the
same sense as Schr\"odinger's relation (\ref{Sinequality}) entails
but is more precise than Heisenberg's relation
(\ref{Rinequality}). According to ({\ref{tauF}), the last
inequality in (\ref{nondissTE}) implies that property $\mean{F}$
cannot change at rates faster than $2\Delta_F\Delta_H/\hbar$.

For dissipative dynamics let us first consider an observable $A$
that commutes with $H$, so that $\com{A}{H}= 0$ while
$\cov{A}{H}\ne 0$; in other words, an observable conserved by the
Hamiltonian term in the dynamical law (\ref{rhodotHM}), but  not
conserved by the dissipative term. Then Equation\ (\ref{exactTE})
yields the equivalent time--energy uncertainty relations
\begin{equation}\label{dissTEA}
\frac{\hbar/2}{\tau_{A}\Delta_H}= a_\tau\,|\rAM|\le a_\tau \ ,
 \end{equation}
\begin{equation}\label{dissTEA2}
\frac{\Boltz\tau }{\tau_{A}\Delta_M}= |\rAM|\le 1 \ .
 \end{equation}

We note that while $\rrAM\le 1$, the value of $a_\tau$ depends on
how $\Delta M(\rho)/\tau$ is defined and, a priori, could well be larger
than unity, in which case there could be some observables $A$ for
which $\tau_{A}\Delta_H\le \hbar/2$. If instead we impose that the
operator function $\Delta M(\rho)/\tau $ is defined in such a way that $a_\tau\le
1$, i.e.,
\begin{equation}\label{taubound}
\tau \ge \hbar\Delta_M\big/2\Boltz\Delta_H \ ,
 \end{equation}
 then we obtain that  even in dissipative
dynamics the usual time--energy uncertainty relations are never
violated by observables $A$ commuting with $H$. In Section \ref{Numerical} we will consider a numerical example for a case with non-constant $\tau$ given by Equation\ \eqref{taubound} with strict equality, for a qualitative comparison with the same case with constant $\tau$.

However, in general, if the dynamics is dissipative
[$\Delta M(\rho)/\tau\ne 0$] there are density operators for which
$|\cFH+a_\tau\,\rFM | >1$ so that $\tau_{F}\Delta_H$ takes a value
less than $\hbar/2$ and thus the usual time--energy uncertainty
relation is violated. The sharpest general time--energy uncertainty
relation that is always satisfied when both Hamiltonian and dissipative
dynamics are active is (proof in Section \ref{Shortest})
\begin{equation}\label{genunc1}
\frac{\hbar^2/4}{\tau^2_{F}\scov{H}{H}} \le 1+a_\tau^2+2a_\tau\cMH
\ ,
\end{equation}
which may also take the equivalent form
\begin{equation}\label{genunc2}
\frac{\tau^2_{F}\scov{H}{H}}{\hbar^2/4}+
\frac{\tau^2_{F}\scov{M}{M}}{\Boltz^2\tau^2(\rho)}+
\frac{\tau^2_{F}\Delta_M\Delta_H\cMH}{\Boltz\tau \,\hbar/4}\ge
1 \ .
\end{equation}

 The upper bound in the rate of change of
 property $
\mean{F}$ becomes
\begin{equation}\label{bound}
\Delta_F\sqrt{\frac{\scov{H}{H}}{\hbar^2/4}+
\frac{\scov{M}{M}}{\Boltz^2\tau }+
\frac{\Delta_M\Delta_H\cMH}{\Boltz\tau \,\hbar/4}}\ .
\end{equation}

As anticipated,  because the dissipative term in Equation 
(\ref{rhodotHM}) implies an additional dynamical mechanism, this
bound (\ref{bound}), valid for the particular nonunitary dynamics
we are considering, is higher than the standard bound valid in
unitary hamiltonian dynamics, given by $2\Delta_F\Delta_H/\hbar$.
For observables commuting with $H$, however, ({\ref{dissTEA2})
provides the sharper general bound $\Delta_F\Delta_M/\Boltz\tau$,
solely due to dissipative dynamics,  which is lower  than
({\ref{bound}).

Because in general $|\cMH|<1$, ({\ref{genunc2}) obviously implies
the less precise relation
\begin{equation}\label{genunc8}
\frac{\hbar^2/4}{\tau^2_{F}\scov{H}{H}} \le (1+a_\tau)^2 \ .
\end{equation}

However, as for the dynamics we discuss in Section \ref{Example},
if the Massieu operator  $\Delta M(\rho)$ is a linear
combination (with coefficients that may depend nonlinearly on
$\rho$) of operators that commute with either  $\rho$ or  $H$,
then it is easy to show that $\cMH=0$. Therefore, in such
important case, ({\ref{genunc2}) becomes
\begin{equation}\label{genunc9}
\frac{\hbar^2/4}{\tau^2_{F}\scov{H}{H}} \le 1+a_\tau^2 \ ,
\end{equation}
clearly sharper than ({\ref{genunc8}).  If in addition
$\Delta M(\rho)/\tau$ is such that  (\ref{taubound}) is satisfied, then (\ref{genunc9})
implies $\tau_{F}\Delta_H \ge \hbar/2\sqrt{2}$.

\section{\label{Entropy}Characteristic Time of the Rate of Entropy Change }

We now consider the entropy  functional $\mean{S}= \Tr(\rho S) =
-\Boltz\Tr (\rho\ln\rho) =-\Boltz \left(\sq \left|\sq
\ln(\sq)^2\right.\right)$ and its rate of change, which using
Equations\ (\ref{rhodotHM}) and (\ref{rateM}) may be written as

\begin{equation}
\label{rateS} {\rm d}\Tr(\rho S)/{\rm d}t =
2\left(\left.\sq S \right|\sq
E(\rho)\right)=\cov{S}{M}\big/\Boltz\tau=
\Delta_S\Delta_M\, \rSM\big/\Boltz\tau\ ,
\end{equation}
where $S$ is  the entropy operator defined as follows 
\begin{equation}
\label{Sdef} S=-\Boltz P_{\Ran\rho}\ln\rho = -\Boltz
\ln(\rho+P_{\Ker\rho} )  \ ,
\end{equation}
where $P_{\Ran\rho}$ and $P_{\Ker\rho}$ are the projection
operators onto the range and kernel of $\rho$. Operator $S$, introduced in \cite{Cimento1,ROMP},
is always well defined for any $\rho\ge\rho^2$, even if some eigenvalues of $\rho$ are zero. It is
the null operator when $\rho^2=\rho$. In models where $S$ is always multiplied by $\rho$ or $\sq$, the operators $P_{\Ran\rho}$ (or  $P_{\Ker\rho}$) in Equation\ \ref{Sdef} could be omitted, because in general $\rho S=-\Boltz\rho\ln\rho$ and $\sq S=-\Boltz\sq\ln\rho$. But, in models of decoherence and composite systems based on the LSEA equation of motion proposed in \cite{Cimento2}, further discussed in \cite{Bregenz09}, and applied for example in \cite{Cano15,Smith16}, their role is important because the LSEA master evolution equation involves the  operators 
\begin{eqnarray}
\label{privateH}&(H)^\J=\Tr_\Jbar [(I_\J\otimes \rho_\Jbar) H]\ ,& \\
\label{privateS}&(S)^\J=\Tr_\Jbar [(I_\J\otimes \rho_\Jbar) S]\ , &
\end{eqnarray}
that we call ``locally perceived overall-system energy operator'' and ``locally perceived overall-system entropy operator,'' respectively, associated with a mean-field-like measure of how the overall-system energy and entropy operators, $H$ and $S$, are ``perceived'' locally within the $J$-th constituent subsystem. The symbol $\Jbar$ denotes the composite of all subsystems except the $\J$-th one. As discussed in full details in \cite{Cimento2,Bregenz09} the  dissipative term in our LSEA master equation points in the direction of the local constrained gradient of the ``locally perceived overall-system entropy'' $\Tr_\J[\rho_\J(S)^\J]$, constrained by the condition of orthogonality with respect to the local gradient of the ``locally perceived overall-system energy'' $\Tr_\J[\rho_\J(H)^\J]$. Operators $(S)^\J$ and, hence,  the LSEA models just mentioned, would not be well defined without  $P_{\Ran\rho}$ (or  $P_{\Ker\rho}$) in Equation\ \eqref{Sdef}.

 Interestingly, the rate of entropy
change, being proportional to the correlation coefficient between
entropy measurements and $M$ measurements, under the assumptions
made so far, may be positive or negative, depending on how $\Delta M(\rho)$ is
defined, i.e., depending on the specifics of the physical model in
which Equation\ (\ref{rhodotHM}) is adopted.

The characteristic time of change of the entropy functional,
defined as
\begin{equation}
\tau_{S}=\Delta_S\big/|{\rm d}\mean{S}/{\rm d}t| \
,\label{tauS}
\end{equation}
gives rise to the following equivalent exact time--energy
uncertainty relations
\begin{equation}
\frac{\hbar/2}{\tau_{S}\Delta_H}=a_\tau\,|\rSM | \le a_\tau \
,\label{TES}
\end{equation}
\begin{equation}\label{genunc3}
\frac{\Boltz\tau }{\tau_{S}\Delta_M}=|\rSM | \le 1\ ,
\end{equation} where $r_{\!S\!M}$  is defined as in
(\ref{rFG}) using operators $\Delta M(\rho)$ and $\Delta
S=S-\mean{S}$. The physical interpretation of~(\ref{genunc3}) is
that the entropy cannot change in time at a rate faster than
$\Delta_S\Delta_M/\Boltz\tau$, as immediately obvious also from
(\ref{rateS}).

We notice from (\ref{TES}) that if the nonequilibrium Massieu operator satisfies condition (\ref{taubound}) then $a_\tau\le
1$ and, therefore, the characteristic time of entropy change,  $\tau_S$,
satisfies the usual uncertainty relation $\tau_{S}\Delta_H\ge
\hbar/2$ and the rate of entropy change cannot exceed
$2\Delta_S\Delta_H/\hbar$.

We conclude this Section by noting that, in general, the equality
in (\ref{TES}) may be used to rewrite
 Relation (\ref{genunc1}) in the form
\begin{equation}\label{genunc4}
\frac{a_\tau}{1+a_\tau} |\rSM|\tau_S\le
\tau_F\frac{\sqrt{1+a^2_\tau+2 a_\tau\cMH}}{1+a_\tau}\le \tau_F\ ,
\end{equation}
where the last inequality follows from $|\cMH|\le 1$. This
relation shows, on one hand, that the entropy change
characteristic time $\tau_S$ is not necessarily the shortest among
the characteristic times $\tau_F$ associated with observables of
the type $\mean{F}=\Tr(\rho F)$ according to the Mandelstam--Tamm
definition (\ref{tauF}). On the other hand, it also shows that the
left-hand side defines a characteristic-time functional
\begin{equation}\label{tauHD}
\tau_{U\! D}=\frac{a_\tau}{1+a_\tau} |\rSM|\tau_S\le \tau_F\ ,
\end{equation}
which constitutes a general lower bound for all $\tau_F$'s, and
may therefore be considered the shortest characteristic time of
 simultaneous unitary$+$dissipative dynamics as described by
Equation (\ref{rhodotHM}). This observation prompts the discussion
in the next section.

\section{\label{Shortest}Shortest Characteristic Times for
Purely-Unitary and Purely-Dissipative Dynamics}

The Mandelstam--Tamm definition (\ref{tauF}) of characteristic times
has been criticized for various reasons (see for example References
\citen{Eberly,Leubner,Bhattacharyya}) mainly related to the fact that
depending on which observable $F$ is investigated, as seen by
inspecting (\ref{nondissTE}), the bound $\tau_F\ge \hbar/2\Delta_H$
may be very poor whenever $\ccFH$ is much smaller than 1.

Therefore, different attempts have been made to define characteristic
times that (1) refer to the quantum system as a whole rather than to
some particular observable, and (2) bound all the particular
$\tau_F$'s from below. Notable examples are the characteristic times
$\tau_{ES}$ and $\tau_{LK}$, respectively defined by Eberly and Singh
\cite{Eberly} and  Leubner and Kiener \cite{Leubner}.

Here, however, we consider the shortest characteristic times that
emerge from the following  geometrical observations. The functional
$\Delta_F$ may be interpreted as the norm of $\sq \Delta F$ (viewed
as a vector in $\Ell (\Hil)$) in the sense that it equals $\sqrt{(\sq
\Delta F |\sq\Delta F)}$, therefore, we may use it to define the
(generally non hermitian) unit norm vector in $\Ell (\Hil)$
\begin{equation}
\label{normalized}  \tilde F_\rho =\sq \Delta F\big/\Delta_F \ .
\end{equation}

As a result, Equation (\ref{rateF2}) may be rewritten in the form
\begin{equation}
 \label{rateFtilde} \frac{1}{\Delta_F} \frac{ {\rm d} \mean{F}}{{\rm d}t}
 =\frac{\Delta_H}{\hbar/2}\,(\tilde F_\rho|i\tilde H_\rho) +
  \frac{\Delta_M}{\Boltz \tau}\,(\tilde F_\rho|\tilde M_\rho)= (\tilde F_\rho|C)\ ,
\end{equation}
where for shorthand we define the  operator
\begin{equation}
\label{defC}  C = i\frac{\Delta_H \tilde
H_\rho}{\hbar/2}+\frac{\Delta_M \tilde M_\rho}{\Boltz\tau}=2\sq
E(\rho) \ ,
\end{equation}
directly related [see Equation (\ref{rateF})] with the evolution
operator function $E(\rho)$ defined in Section \ref{Structure}, which
determines the rates of change of all linear
 functionals of the state operator $\rho$, i.e., all observables of
 the linear type $\Tr(\rho F)$, by  its
   projection onto the respective directions $\tilde F_\rho$.

Each characteristic time $\tau_F$ can now be written as
\begin{equation}
\label{tauF2}  \tau_F=\Delta_F\big/|{\rm d} \mean{F}/dt|=1\big/|(\tilde
F_\rho|C)| \   .
\end{equation}

Because $\tilde F_\rho$ is unit norm, $|(\tilde F_\rho|C)|$ is
bounded by the value attained for an operator $\tilde F_\rho$ that
has the same `direction' in $\Ell (\Hil)$ as operator $C$, i.e.,
for
\begin{equation}
\label{frho} \tilde F_\rho = \pm C\Big/\sqrt{(C|C)} \   ,
\end{equation}
in which case $|(\tilde
F_\rho|C)|=\sqrt{(C|C)}=\sqrt{\Tr(C^\dagger C)}$. Thus we conclude
that, for any, $F$,
\begin{equation}
\label{ineqCF}1\Big/\sqrt{(C|C)}\le \tau_F \   ,
\end{equation}
and, therefore, we introduce the shortest characteristic time for
the combined unitary$+$dissipative dynamics described by Equation
(\ref{rhodotHM}),
\begin{equation}
\label{tauUD}\tau_{U\! D}=1\Big/\sqrt{(C|C)} \   ,
\end{equation}
which binds from below all $\tau_F$'s. From (\ref{defC}) and
(\ref{ineqCF}), and the identities $(i\tilde H_\rho |i\tilde
H_\rho)=(\tilde M_\rho |\tilde M_\rho)=1$ and $(i\tilde H_\rho
|\tilde M_\rho)=(\tilde M_\rho |i\tilde H_\rho)=\cMH$ we obtain

\begin{equation}
\begin{array}{lll}
\label{ineqUD} \frac{1}{\tau_F^2}\le\frac{1}{\tau_{U\!
D}^2}&=&(C|C)= \frac{\scov{H}{H}}{\hbar^2/4}+
\frac{\scov{M}{M}}{\Boltz^2\tau^2(\rho)}+
\frac{\Delta_M\Delta_H\cMH}{\Boltz\tau \,\hbar/4}\\
&=&\frac{\scov{H}{H}}{\hbar^2/4}(1+a_\tau^2+2\,a_\tau\cMH) \ ,
\end{array}
\end{equation}
which proves relations (\ref{genunc1}) and (\ref{genunc2}).

For nondissipative (purely Hamiltonian, unitary) dynamics the same
reasoning (or substitution of $\tau=\infty$, $a_\tau=0$ in the
above relations) leads to the definition of the shortest
characteristic time of unitary dynamics
\begin{equation}
\label{deftauU}\tau_U=\hbar\big/2\Delta_H \   ,
\end{equation}
with which the usual time--energy relation reduces to
\begin{equation} \label{genunc11}\tau_F\ge \tau_U \   .
\end{equation}

Its physical meaning is that when the energy dispersion (or
uncertainty or spread) $\Delta_H$ is small, $\tau_U$ is large and
$\tau_F$ must be larger for all observables $F$, therefore, the
mean values of all properties change slowly \cite{Pfeifer93_1,Pfeifer93_2,Pfeifer95}, i.e.,
the state $\rho$ has a long lifetime. In other words,  states with a
small energy spread cannot change rapidly with time. Conversely, states that
change rapidly due to unitary dynamics, necessarily have a large
energy spread.

Another interesting extreme case obtains from Equation (\ref{rhodotHM})
when $\Delta M(\rho)$ is such that the condition $[\rho,H]=0$
implies $[\Delta M(\rho),H]=0$ for any $\rho$, as for the
steepest-entropy-ascent dynamics discussed in Sections
\ref{Example} and \ref{Numerical}. In this case, it is easy to see
that if the state operator $\rho$ commutes with $H$ at one instant
of time then it commutes with $H$ at all times and, therefore, the
entire time evolution is purely dissipative. Then, the reasoning
above leads to the definition of the shortest characteristic time of purely dissipative evolution
\begin{equation}
\label{deftauD}\tau_D=\Boltz\tau/\Delta_M \   .
\end{equation}

It is noteworthy that $\tau_D$ can be viewed as the characteristic
time associated not with the (generally nonlinear) Massieu
functional $\mean{M}=\Tr(\rho M(\rho))$ but with the linear
functional $\mean{A}=\Tr(\rho A)$ corresponding to the
time-independent operator $A$ which at time $t$ happens to
coincide with $M(\rho(t))$.

For purely dissipative dynamics, the bound
$\tau_F\ge\tau_D=\Boltz\tau/\Delta_M$ implies that when
$\Delta_M/\Boltz\tau$, i.e., the ratio between the uncertainty in our
generalized nonequilibrium Massieu observable represented by operator $M$ and the intrinsic dissipation
time $\tau$, is small, then $\tau_D$ is large and $\tau_F$
must be larger for all observables $F$, therefore, the state
$\rho$ has a long lifetime. This may be a desirable feature in quantum computing applications where the interest is in engineering states $\rho$ that preserve the entanglement of component subsystems. Conversely, if some observable changes
rapidly, $\tau_F$ is small and since $\tau_D$ must be smaller, we
conclude that the spread $\Delta_M$ (more precisely, the ratio
$\Delta_M/\Boltz\tau $) must be large.

In terms of $\tau_U$ and $\tau_D$ we can rewrite (\ref{ataudef}),
(\ref{genunc3}) and (\ref{ineqUD}) as
\begin{equation}
a_\tau=\tau_U/\tau_D \   ,
\end{equation}
\begin{equation}\label{tauSD}
\frac{1}{\tau_S}=\frac{|\rSM|}{\tau_D}\le \frac{1}{\tau_D} \   ,
\end{equation}
\begin{eqnarray}
\label{ineqUD2}
\frac{1}{\tau_F^2}&=&\left(\frac{\cFH}{\tau_U}+\frac{\rFM}{\tau_D}\right)^2
\nonumber\\ &\le&\frac{1}{\tau_{U\!
D}^2}=\frac{1}{\tau_U^2}+\frac{1}{\tau_D^2}
+\frac{2\,\cMH}{\tau_U\tau_D} \\
&\le&\left(\frac{1}{\tau_U}+\frac{1}{\tau_D}\right)^2  .\nonumber
\end{eqnarray}

Equation (\ref{tauSD}) implies that the entropy cannot change rapidly
with time  if the ratio $\Delta_M(\rho)/\Boltz\tau $ is not
large. The first equality in (\ref{ineqUD2}) follows from $(\tilde
F_\rho |i\tilde H_\rho)=\cFH$ and $(\tilde F_\rho |\tilde
M_\rho)=\rFM$, which also imply that  Equation (\ref{rateFtilde}) may
take the form
\begin{equation}
 \label{rateFtilde2}   \frac{{\rm d} \mean{F}}{{\rm d}t}
  =\Delta_F\left(\frac{\cFH}{\tau_U} +
  \frac{\rFM}{\tau_D}\right)
 \ ,
\end{equation}
and operator $C$ defined in (\ref{defC}) takes also the forms
\begin{equation}
C= i\frac{\tilde H_\rho}{\tau_U} +
 \frac{\tilde M_\rho}{\tau_D}= i\frac{\sq\Delta H}{\Delta_H\tau_U}+\frac{\sq\Delta
 M}{\Delta_M\tau_D} \ ,
 \end{equation} and its norm is
 $\sqrt{1/\tau_U^2+1/\tau_D^2+2\cMH/\tau_U\tau_D}$.

 Similarly, the rate of entropy change (\ref{rateS}) takes the
 form
\begin{equation} \label{rateS2} \frac{{\rm d}\mean{S}}{{\rm d}t} =
\frac{\Delta_S}{\tau_D}\,(\tilde S_\rho\,|\,\tilde
M_\rho)=\frac{\Delta_S\, \rSM}{\tau_D}
\end{equation}
which, because $|\rSM |\le 1$, implies the bounds [equivalent to
(\ref{genunc3}) and (\ref{tauSD})],
\begin{equation} \label{rateSbound} -\frac{\Delta_S}{\tau_D}\le
\frac{{\rm d}\mean{S}}{{\rm d}t} \le \frac{\Delta_S}{\tau_D} \ .
\end{equation}

\section{\label{Occupation}Occupation Probabilities}

An important class of observables for a quantum system are those
associated with the projection operators. For example,
 for pure states evolving unitarily, the mean value
 $\mean{P}=\Tr(\rho(t)
P)$ where $P=|\phi_0\rangle\langle\phi_0|=\rho(0)$ represents the
survival probability of the initial state, and is related to
several notions of lifetimes \cite{Pfeifer93_1,Pfeifer93_2,Pfeifer95}.

We do not restrict our attention to pure states, and we discuss
 first  results that hold for any projector $P$ associated with a
 yes/no type of measurement. Let $P=P^\dagger=P^2$ be an
 orthogonal projector onto the $g$-dimensional subspace $P\Hil$ of
 $\Hil$. Clearly, $g=\Tr(P)$, the variance
 $\cov{P}{P}=p\,(1-p)$ where $p=\mean{P}=\Tr(\rho
P)$ denotes the mean value and represents the probability in state
$\rho$ of obtaining a `yes'  result upon measuring the associated
observable. The characteristic time of the rate of change of
this occupation probability is defined according to (\ref{tauF})
by

\begin{equation}
\begin{array}{lll}
\label{tauP}\frac{1}{\tau_{P}}&=&\frac{|{\rm d}p/{\rm
d}t|}{\sqrt{p\,(1-p)}}=2\left|\frac{\rm d}{{\rm
d}t}\arccos(\sqrt{p})\right|\\
&=& 2\left|\frac{\rm
d}{{\rm d}t}\arcsin(\sqrt{p})\right|\le\frac{1}{\tau_{U\! D}}\ ,
\end{array}
\end{equation}
where the inequality follows from (\ref{ineqUD}). Therefore,
\begin{equation}
\label{cospn}-\frac{1}{2\tau_{U\! D}}\le\frac{\rm d}{{\rm
d}t}\arccos(\sqrt{p})\le \frac{1}{2\tau_{U\! D}} \ ,
\end{equation}
or, over any finite time interval of any time history $p(t)$,
\begin{equation}
\label{cosfinite}\left|\arccos(\sqrt{p(t_2)})-\arccos(\sqrt{p(t_1)})\right|\le
\left|\int_{t_1}^{t_2}{\frac{{\rm d}t'}{2\tau_{U\! D}(t')}}\right|
\ .
\end{equation}

This result generalizes the results on lifetimes  obtained in
\cite{Bhattacharyya} where the focus is restricted to full quantum
decay [$p(\infty)\approx 0$] of an initially fully populated state
[$p(0)\approx 1$] and $\tau_U$ (here $\tau_{U\! D}$) is assumed
constant during the time interval. It is also directly related to
some of the results in \cite{Pfeifer93_1,Pfeifer93_2,Pfeifer95}, where a number of
additional inequalities and bounds on lifetimes are obtained for
unitary dynamics, and may be straightforwardly generalized to the
class of simultaneous unitary/dissipative dynamics  described by
our Equation (\ref{rhodotHM}).

Because $p\,(1-p)$ attains its maximum value when $p=1/2$, we also
have the inequality
\begin{equation}
\label{dpndt}\left|\frac{{\rm d}p}{{\rm d}t}\right|\le
\frac{1}{2\tau_{U\! D}} \ .
\end{equation}
which, analogously to what noted in  \cite{Bhattacharyya}, implies
that no full decay nor full population can occur within a time
$2\tau_{U\! D}$, so that this time may be interpreted as a limit
to the degree of instability of a quantum state.

Next, we focus on the projectors onto the eigenspaces of the
Hamiltonian operator $H$, assumed time-independent. Let us write
its spectral expansion as $H=\sum_n e_n\Pen$ where $e_n$ is the
$n$-th eigenvalue and $\Pen$ the projector onto the corresponding
eigenspace. Clearly, $H\Pen =e_n\Pen$, $\Pen\Pem=\delta_{nm}\Pen$,
 $g_n=\Tr(\Pen )$ is the degeneracy of eigenvalue
$e_n$, $p_n=\mean{\Pen }=\Tr(\rho \Pen )$ the occupation
probability of energy level $e_n$, $\cov{\Pen }{\Pem
}=p_n\,(\delta_{nm}-p_m)$ the covariance of pairs of occupations,
and $\cov{\Pen }{\Pen }=p_n\,(1-p_n)$ the variance or fluctuation
of the $n$-th occupation. Because $[\Pen ,H]=0$, $\cPenH=0$ and by~(\ref{rateFtilde2}) we have
\begin{equation} \label{ratePen}   \frac{{\rm d} p_n}{{\rm d}t}
  =\Delta_{\Pen}
  \frac{\rPenM}{\tau_D}
 \ ,\end{equation}
 and the corresponding characteristic time is
 \begin{equation} \label{teuPen}   \frac{1}{\tau_\Pen}
  =
  \frac{|\rPenM |}{\tau_D} \le \frac{1}{\tau_D}
 \ .\end{equation}

Energy level occupation probabilities $p_n$ are used in Section
\ref{Numerical} for numerical illustration/validation of
inequalities (\ref{teuPen}) within the steepest-entropy-ascent
dynamical model outlined in the next Section.

\section{\label{Example}Example. Steepest-Entropy-Ascent Master Equation for
Conservative Dissipative Dynamics}

So far we have not assumed an explicit form of the operator $\Delta M(\rho)$
except for the condition   that it maintains $\rho$ unit
trace (\eqref{preservetrace1} or \eqref{rateM}) and nonnnegative definite. In this section, we illustrate the above results  by
further assuming a particular form of steepest-entropy-ascent, conservative dissipative
dynamics. For our generalized
nonequilibrium Massieu operator we assume the expression
\begin{equation}
\label{Msteepest} \Delta M(\rho)=\Delta S- \Delta
H'(\rho)/\theta(\rho) \ ,
\end{equation}
where $S$ is the entropy operator defined in Equation (\ref{Sdef}),
\begin{equation}
\label{defH'} \Delta H'(\rho) = \Delta H - \nnu(\rho)\cdot
\Delta\N\ ,
\end{equation}
$H$ is the
Hamiltonian operator, $\N=\{N_1,\dots,N_r\}$  a (possibly empty) set of
operators  commuting with $H$   that we call non-Hamiltonian
generators of the motion (for example, the number-of-particles
operators or a subset of them, or the momentum component operators
for a free particle) and that must be such that operators $\sq\Delta H$ and $\sq
\Delta \N$ are linearly independent, and -- most importantly -- $\theta(\rho)$ and
$\nnu(\rho)=\{\nu_1(\rho),\dots,\nu_r(\rho)\}$ are a set of real
functionals defined for each $\rho$ by the solution of the
following system of linear equations
\begin{eqnarray}
\label{constraint1}
\cov{S}{H}\,\theta+\sum_{i=1}^r\cov{N_i}{H}\,\nu_i&=& \cov{H}{H}\
,\\ \label{constraint2}
\cov{S}{N_j}\,\theta+\sum_{i=1}^r\cov{N_i}{N_j}\,\nu_i&=&
\cov{H}{N_j} \ ,
\end{eqnarray}
which warrant the conditions that $\cov{H}{M}=0$ and
$\cov{N_j}{M}=0$, and hence that the mean values $\Tr(\rho H)$ and
$\Tr(\rho \N)$ are maintained time invariant by the dissipative
term of the resulting SEA master equation [Equation\ (\ref{rhodotHM}) together with Equations\ (\ref{Msteepest})--(\ref{constraint2})].

As a result, our assumption may be rewritten as follows
\begin{equation}\label{deltaM}\Delta M(\rho)=M(\rho)-I\Tr[\rho
M(\rho)]\end{equation} where $I$ is the identity and the nonequilibrium Massieu operator $M(\rho)$ is
the following nonlinear function  of $\rho$
\begin{equation}
\label{Massieu} M(\rho)=  S(\rho)-\frac{ H}{\theta(\rho)}+
\frac{\nnu(\rho)\cdot \N}{\theta(\rho)} \ ,
\end{equation}
and we note that  at a thermodynamic equilibrium (Gibbs) state,
\begin{equation}
\label{rhoe} \rho_{\rm e}=  \frac{1}{Z}\exp\left(-\frac{H-\mmu_{\rm e}\cdot \N}{T_{\rm e}}\right)\ ,
\end{equation}
its mean
value belongs to the family of entropic characteristic functions
introduced by Massieu \cite{Massieu}, i.e.,
\begin{equation}
\label{MassieuEq} \langle M\rangle_{\rm e}= \langle S\rangle_{\rm
	e}-\frac{\langle H\rangle_{\rm e}}{T_{\rm e}}+ \frac{\mmu_{\rm e}\cdot
	\langle\N\rangle_{\rm e}}{T_{\rm e}} \ ,
\end{equation}
where $\langle S\rangle_{\rm e}$, $\langle H\rangle_{\rm e}$,
$\langle\N\rangle_{\rm e}$, $T_{\rm e}=\theta(\rho_{\rm e})$ and $\mmu_{\rm e}=\nnu(\rho_{\rm e})$ are the (grand
canonical) equilibrium entropy, energy, amounts of constituents,
temperature and chemical potentials, respectively.

Notice 
that operator $M$, its eigenvalues and its mean value $\Tr(\rho M)$ for a given state $\rho$, that we first termed ``nonequilibrium Massieu operator''  in  References \cite{PRE06,IJQT07,ROMP}, differ substantially from  the ``nonequilibrium Massieu potentials'' defined recently in References \cite{Rao18NJP,Rao18Entropy}. Their nonequilibrium Massieu construct is defined by the difference between the entropy and a linear combination of the conserved properties, with coefficients that are weighted averages of the fixed temperatures and other entropic potentials of the reservoirs interacting with the system. In our nonequilibrium Massieu construct, instead, the coefficients $\theta$ and $\nnu$ of the linear combination are truly nonequilibrium functionals of the state $\rho$, that evolve in time with $\rho$, and that only when the system has relaxed to equilibrium can be identified with the inverse temperature $1/T$ of the system and the  entropic potentials $-\mmu/T$ of the other conserved properties.

The non-Hamiltonian generators of the motion represent the other conserved properties of the system, however,
this condition may be relaxed in the framework of a resource theory of a quantum thermodynamic subsystem that, via the Hamiltonian part of the master equation, exchanges with other systems or a thermal bath  some non-commuting quantities or ``charges'', as recently envisioned in Reference \cite{Nicole16}.

Operators $\sq\Delta H'$ and $\sq\Delta M$ are always orthogonal to each other,
in the sense that $\cov{M}{H'}=0$ for every $\rho$. It follows
that, in general, $\cov{S}{H'}=\cov{H'}{H'}/\theta$,
\begin{equation}
\label{covSM} \cov{S}{M}= \cov{M}{M}=
\cov{S}{S}-\frac{\cov{H'}{H'}}{\theta^2(\rho)}\ge 0 \ ,
\end{equation}
and hence the rate of entropy generation (\ref{rateS}) is always
strictly positive except for $\cov{M}{M}=0$ (which occurs iff
$\sq\Delta M=0$), i.e., for$ \sqnd\Delta S_{\rm nd}=(\sqnd\Delta H
- \mmu_{\rm nd}\cdot\sqnd\Delta\N)/T_{\rm nd}$, for some real
scalars $T_{\rm nd}$ and $\mmu_{\rm nd}$, that is, for
density operators (that we call non-dissipative \cite{Cimento1,PRE06,IJQT07,ROMP}) of the following Gibbs (or partially Gibbs, if $B\ne I$) form
\begin{equation}
\label{rhonondiss} \rho_{\rm nd}=\frac{B\exp[-( H - \mmu_{\rm
nd}\cdot\N)/\Boltz T_{\rm nd}]B}{\Tr B\exp[-( H - \mmu_{\rm
nd}\cdot\N)/\Boltz T_{\rm nd}]} \ ,
\end{equation}
where $B$ is any projection operator on $\Hil$ ($B^2=B$).

The nonlinear functional
\begin{equation}\label{theta}
\theta(\rho)=\frac{\scov{H'}{H'}}{\scov{S}{H'}}
=\frac{\Delta_{H'}}{\Delta_S\,\rSHprime}
\end{equation}
 may be interpreted in this
framework as a natural generalization to nonequilibrium of the
 temperature, at least insofar as for $t\rightarrow +\infty
$, while the state operator $\rho(t)$ approaches a non-dissipative
operator of form  (\ref{rhonondiss}), $\theta(\rho(t))$ approaches
smoothly the temperature $T_{\rm nd}$ of the non-dissipative thermodynamic equilibrium (stable, if $B= I$, or unstable, if $B\ne I$) or of the unstable limit cycle (if $[B,H]\ne 0$), and $-\nnu(\rho(t))/\theta(\rho(t))$ approach
smoothly the corresponding  entropic potentials $-\mmu_{\rm nd}/T_{\rm nd}$.

Because here we assumed that $H$ always commutes with $M$,  $\cMH=0$ and $(\tilde
M|i\tilde H)=0$, which means that $\sq \Delta M(\rho)$ is always
orthogonal to $i\sq \Delta H$. This reflects the fact that the direction of steepest-entropy-ascent   is
orthogonal to the (constant entropy) orbits that characterize
purely
 Hamiltonian (unitary) motion (which maintains the entropy  constant by
 keeping  invariant each eigenvalue of~$\rho$). 
 
 Here, for simplicity, we have assumed that dissipation pulls the state in  the direction of steepest-entropy-ascent with respect to the uniform Fisher--Rao metric (see \cite{ROMP}). However, we have discussed elsewhere (see \cite{PRE14,PRE15}) that, in general, a most important and characterizing feature of the nonequilibrium states of a system is the metric with respect to which the system identifies the direction of steepest-entropy-ascent. In most cases, it is a non-uniform metric, such as for a material with a nonisotropic thermal conductivity or, in the quantum framework, for a spin system in a magnetic field that near equilibrium obeys the Bloch equations \cite{Bedeaux91} with different relaxation times along the field and normal to the field.

Inequality (\ref{covSM}), which follows from $\rrSM\le 1$, implies
that
    $\scov{M}{M}\le \scov{S}{S}
  $ and $0\le\rSM=\Delta_M/\Delta_S\le 1$ or, equivalently,
\begin{equation}
\label{deftauG}\tau_K=\Boltz\tau/\Delta_S\le \tau_D \   ,
\end{equation}
where for convenience we  define the
   characteristic time $\tau_K$, which is simply related to the entropy uncertainty, but
   cannot be attained by any rate of change, being shorter than~$\tau_D$.
In addition, we have the~identities
\begin{equation}
\label{rrSM}
\rrSM=\frac{\scov{M}{M}}{\scov{S}{S}}=\frac{\tau_K^2}{\tau_D^2}
=\frac{\tau_K}{\tau_S}
=1-\frac{\scov{H'}{H'}}{\theta^2\scov{S}{S}}=1-\rrSHprime \ ,
\end{equation}
and, from $\rrSHprime\le 1$, the bounds
\begin{equation}
\label{thetabound} |\theta| \ge \frac{\Delta_{H'}}{\Delta_S} \quad
{\rm or}\quad
-\frac{\Delta_S}{\Delta_{H'}}\le\frac{1}{\theta}\le\frac{\Delta_S}{\Delta_{H'}}
\ ,
\end{equation}
where the equality $ |\theta| =\Delta_{H'}/\Delta_S$ holds when
and only when the state is non-dissipative [Equation~(\ref{rhonondiss})].  Additional bounds on our generalized
nonequilibrium temperature $\theta$ obtain by combining
(\ref{rrSM}) with the inequality $4\rrSM(1-\rrSM)\le 1$ (which
clearly holds because $ \rrSM\le 1$), to obtain
$4\rrSM\rrSHprime\le 1$ and,~therefore,
\begin{equation}
\label{thetabound2}
\frac{2\Delta_M\Delta_{H'}}{|\theta|\scov{S}{S}}\le 1 \quad {\rm
or}\quad
-\frac{\scov{S}{S}}{2\Delta_M\Delta_{H'}}\le\frac{1}{\theta}\le\frac{\scov{S}{S}}{2\Delta_M\Delta_{H'}}
\ .
\end{equation}

At equilibrium,  $\Delta_M=0$ and (\ref{thetabound2}) implies no
actual bound on $\theta$, but in nonequilibrium states bounds
(\ref{thetabound2}) may be tighter than (\ref{thetabound}), as
illustrated by the numerical example in Section \ref{Numerical}.

   Notice that whereas in steepest-entropy-ascent dynamics
   $\tau_K$ is always shorter than $\tau_D$ and
   obeys
   the identity \begin{equation}\label{identity}\tau_S\tau_K=\tau_D^2\ ,\end{equation}
 in general it is not necessarily shorter than $\tau_D$ and
obeys the identity
\begin{equation}\label{identitygen}\frac{\Delta_M}{\Delta_S}\frac{\tau_D^2}{\tau_S\tau_K}=|\rSM|\ .\end{equation}

In summary, we conclude that
  within steepest-entropy-ascent, conservative dissipative quantum
 dynamics, the general uncertainty relations
 (\ref{genunc2}), (\ref{TES}) and (\ref{genunc3})
 that constitute the main results of this paper,
yield the
 time--energy/time-Massieu uncertainty relation
 \begin{equation}\label{genunc6M}
\left(\frac{\tau_{F}\Delta_H}{\hbar/2}\right)^2+
\left(\frac{\tau_{F}\Delta_M}{\Boltz\tau }\right)^2
 \ge 1 \quad {\rm or}\quad
   \frac{\tau^2_{F}}{\tau^2_{U}}+\frac{\tau^2_{F}}{\tau^2_{D}} \ge 1 \ ,
\end{equation}
which implies the  interesting
 time--energy and time--entropy uncertainty relation
 \begin{equation}\label{genunc6}
\left(\frac{\tau_{F}\Delta_H}{\hbar/2}\right)^2+
\left(\frac{\tau_{F}\Delta_S}{\Boltz\tau }\right)^2 \ge 1
\quad {\rm or}\quad
   \frac{\tau^2_{F}}{\tau^2_{U}}+\frac{\tau^2_{F}}{\tau^2_{K}} \ge 1 \ ,
\end{equation}
 and the  time--entropy  uncertainty relation
\begin{equation}\label{genunc5}
\frac{\tau_K}{\tau_{S}}=\frac{\Boltz\tau }{\tau_{S}\Delta_S}=\rrSM
\le \rSM \le 1 \ ,
\end{equation}
which implies that the rate of entropy generation never exceeds
$\scov{S}{S}/\Boltz\tau$, i.e.,
\begin{equation}\label{genunc7}
\frac{{\rm d}\mean{S}}{{\rm d}t}=-\Boltz\frac{{\rm d}}{{\rm
d}t}\Tr (\rho \ln\rho) =\frac{\scov{M}{M}}{\Boltz\tau}
\le\frac{\Delta_S\Delta_M}{\Boltz\tau}
\le\frac{\scov{S}{S}}{\Boltz\tau}\ .
\end{equation}

If in addition the dynamics is purely dissipative, such as along a
trajectory $\rho(t)$ that commutes with $H$ for every $t$, then
(\ref{genunc6}) may be replaced by the time--entropy uncertainty
relation
 \begin{equation}\label{genunc12}
\frac{\tau_K}{\tau_{F}}=\frac{\Boltz\tau }{\tau_{F}\Delta_S}
\le 1 \ .
\end{equation}

As shown in References \citen{Cimento1,ROMP}, the dissipative dynamics
generated by Equation\ (\ref{rhodotHM}) with $\Delta M(\rho)$ as just
defined and a time-independent Hamiltonian $H$: (i) maintains
$\rho(t)\ge \rho^2(t)$ at all times, both forward and backward in
time for any initial density operator $\rho(0)$ (see also
\cite{Gheorghiu1,Gheorghiu2}); (ii) maintains the cardinality of $\rho(t)$
invariant; (iii) entails that the entropy functional is an
$S$-function in the sense defined in \cite{Lyapunov} and therefore
that maximal entropy density operators (Gibbs states) obtained from
(\ref{rhonondiss}) with $B=I$ are the only
 equilibrium states of the dynamics that are stable with respect
 to perturbations that do not alter the mean values of the energy
 and the other time invariants (if any): this theorem of the
 dynamics  coincides with the Hatsopoulos-Keenan statement of
 the second law of thermodynamics~\cite{Book};
(iv)  entails Onsager reciprocity in the sense defined in
 \cite{Onsager}; (v) can be derived from a variational principle
 \cite{Gheorghiu1,Gheorghiu2}, equivalent to our steepest-entropy-ascent
 geometrical construction,
 by maximizing the entropy generation rate subject to the $\Tr(\rho)$,
  $\Tr(\rho H)$, and $\Tr(\rho\N)$ conservation  constraints and
  the additional constraint   $(\sq E|\sq E)=c(\rho)$.
  
  Operator
  $\sq E$ is a `vector' in
  $\Ell(\Hil)$ and determines through its scalar product with $\sq
  F$ and $\sq S$ [Equations\ (\ref{rateF}) and (\ref{rateS})] the rates
  of change of $\Tr(\rho F)$ and $\Tr(\rho S)$, respectively. From
  (\ref{rateS}) and the Schwarz inequality $(\sq S|\sq E)^2\le (\sq
  S|\sq S)(\sq E|\sq E)$, we see that for a given $\rho$, among all
  vectors $\sq E$ with  given norm $(\sq E|\sq
  E)=c(\rho)$, the one  maximizing $(\sq S|\sq E)$ has the same
  direction as $\sq S$. In general, along such direction $\Tr(\rho
  H)$ and $\Tr(\rho\N )$ are  not conserved because $\sq S$ is not
  always  orthogonal to $\sq H$ and $\sq \N$. Instead, dynamics
  along the direction of steepest-entropy-ascent compatible with
  such conservation requirements, as first postulated and formulated
  in \cite{thesis,Cimento1,ROMP}, obtains when $\sq E$ has the direction of the
  component of $\sq S$ orthogonal to $\sq H$ and $\sq \N$. This is
  precisely how $\Delta M(\rho)$ is defined through Equations\ 
  (\ref{Msteepest})--(\ref{constraint2}). See also Reference~\cite{Bregenz09}.

We finally note that assuming in Equation (\ref{rhodotHM}) a $\Delta M(\rho)/\tau$ that satisfied Equation\ (\ref{taubound}) with strict
equality, we obtain the most dissipative (maximal entropy
generation rate) dynamics in which the entropic
characteristic time $\tau_S$ (Equation\ (\ref{tauS})) is always
compatible with the time--energy uncertainty relation
$\tau_S\Delta_H\ge \hbar/2$ and the rate of entropy generation is
always given by $2\Delta_M\Delta_H/\hbar$.

The physical meaning of relations (\ref{genunc2}), (\ref{TES}),
(\ref{genunc3}), (\ref{genunc6}), (\ref{genunc5}) are worth
further investigations and experimental validation in specific
contexts in which the dissipative behavior is correctly modeled by
a dynamical law of form (\ref{rhodotHM}), possibly with $\Delta
M(\rho)/\tau$ of form (\ref{Msteepest}). One such context may be the
currently debated so-called ``fluctuation theorems'' \cite{Evans94,Crooks99,Dewar03,Cohen03}
whereby fluctuations and, hence, uncertainties are measured on a
microscopic system (optically trapped colloidal particle
\cite{Wang,Nieuwenhuizen}, electrical resistor \cite{Garnier}) driven at steady
state (off thermodynamic equilibrium) by means of a work
interaction, while a heat interaction (with a bath) removes the
entropy being generated by irreversibility. Another such context
may be that of pion-nucleus scattering, where available
experimental data have recently allowed partial validation
\cite{Ion} of ``entropic'' uncertainty relations \cite{Deutsch,Partovi,Maassen}.
Yet another is within the model we propose in Reference  \citen{PRE06}
for the description of  the irreversible time evolution of  a
perturbed, isolated, physical system during relaxation toward
thermodynamic equilibrium by spontaneous internal rearrangement of
the occupation probabilities. We pursue this example in the next
section.

\section{\label{Numerical}Numerical Results for Relaxation within
a Single $N$-Level Qudit or a One-Particle Model of a Dilute Boltzmann
Gas of $N$-Level Particles}

To illustrate the time dependence of the uncertainty relations
derived in this paper, we consider an isolated, closed system
composed of noninteracting identical particles with
single-particle eigenstates with energies $e_i$ for $i=1$,
2,\dots, $N$,  where $N$ is assumed finite for simplicity and the
$e_i$'s are repeated in case of degeneracy, and we restrict our
attention to the class of dilute-Boltzmann-gas states in which the
particles are independently distributed among the $N$ (possibly
degenerate) one-particle energy eigenstates. This model is
introduced in Reference \citen{PRE06}, where we assume an equation of
 form (\ref{rhodotHM}) with $\Delta M(\rho)$ given by (\ref{Msteepest}) with the further simplification that
$\Delta H'(\rho) = \Delta H$ so that our generalized
nonequilibrium Massieu operator is simply
\begin{equation}
\label{MsteepestH} M(\rho)= S-  H/\theta(\rho) \ ,
\end{equation}
and, therefore,
\begin{equation}
\label{deltaMsteepestH} \Delta M(\rho)=\Delta S- \Delta
H/\theta(\rho) \ .
\end{equation}

 For simplicity and illustrative purposes, we focus on purely dissipative
 dynamics by considering a particular
trajectory $\rho(t)$ that commutes with $H$ at all times $t$,
assuming that $H$ is
 time independent and has a nondegenerate
spectrum. As a result, the energy-level occupation probabilities
$p_n$ coincide with the eigenvalues of $\rho$, and the dynamical
equation reduces to the simple form \cite{PRE06}
\begin{equation}
\label{pdot} \frac{{\rm d}p_n}{{\rm
d}t}=-\frac{1}{\tau}\left[p_n\ln p_n +
p_n\frac{\mean{S}}{\Boltz}+p_n\frac{e_n-\mean{H}}{\Boltz\theta}\right]
\ ,
\end{equation}
where
\begin{eqnarray}
\label{pdotdefs} \mean{S}&=&-\Boltz\sum_n p_n\ln p_n \ ,\\
\mean{H}&=&\sum_n p_ne_n \ ,\\ \theta&=&\scov{H}{H}/\scov{H}{S} \
,\\ \scov{H}{H}&=&\sum_n p_ne^2_n-\mean{H}^2 \ ,\\
\scov{H}{S}&=&-\Boltz\sum_n p_ne_n\ln p_n -\mean{H}\mean{S} \ .
\end{eqnarray}

The same model describes relaxation to the Gibbs state of an $N$-level qudit with time independent Hamiltonian $H$ from arbitrary initial states $\rho(0)$ that commute with $H$.

To obtain the plots in Figures \ref{Figure1} and \ref{Figure2}, that illustrate the main
inequalities derived in this paper for a sample trajectory, we
consider  an initial state with cardinality equal to 4, with
nonzero occupation probabilities only for the four energy levels
$e_1=0$, $e_2=u/3$, $e_3=2u/3$, and $e_4=u$, and with mean energy
$\mean{H}=2u/5$ ($u$ is arbitrary, with units of energy).
Moreover, as done in \cite{PRE06}, we select an  initial state
$\rho(0)$ at time $t=0$ such that the resulting trajectory
$\rho(t)$ passes in the neighborhood of the partially canonical
nondissipative state $\rho^{\rm ft}_{\rm nd}$ that has  nonzero
occupation probabilities only for the three energy levels $e_1$,
$e_2$, and $e_4$, and mean energy $\mean{H}=2u/5$ ($p^{\rm
ft}_{{\rm nd}1}=0.3725$, $ p^{\rm ft}_{{\rm nd}2}=0.3412$, $p^{\rm
ft}_{{\rm nd}3}=0$, $p^{\rm ft}_{{\rm nd}4}=0.2863$, $\theta^{\rm
ft}_{\rm nd} =3.796\, u/\Boltz$ ). As shown in Figure \ref{Figure1}, during
the first part of the trajectory, this nondissipative state
appears as an attractor, an approximate or `false target'
equilibrium state; when the trajectory gets close to this state,
the evolution slows down, the entropy generation drops almost to
zero and the value of $\theta$ gets very close ($3.767\,
u/\Boltz$) to that of $\theta^{\rm ft}_{\rm nd}$; however
eventually the small, but nonzero initial occupation of level
$e_3$ builds up and a new rapid rearrangement of the occupation
probabilities takes place, and finally drives the system toward
the maximal entropy state $\rho^{\rm pe}_{\rm nd}$ with energy
$\mean{H}=2u/5$ and all four active levels occupied, with
 canonical (Gibbs) distribution $p^{\rm e}_{{\rm nd}1}=0.3474$,
$ p^{\rm e}_{{\rm nd}2}=0.2722$, $p^{\rm e}_{{\rm nd}3}=0.2133$,
$p^{\rm e}_{{\rm nd}4}=0.1671$, and  characterized by the equilibrium temperature
$T_{\rm e} =1.366\, u/\Boltz$.

The trajectory is computed by  integrating Equation\ (\ref{pdot})
numerically, both forward and backward in time,
 starting from the chosen initial state $\rho(0)$, and assuming
 for Figures \ref{Figure1}a and \ref{Figure2}a
 that the dissipation time $\tau$ is a constant, and for Figures \ref{Figure1}b
 and \ref{Figure2}b that it is given by (\ref{taubound}) with strict equality
 ($a_\tau=1$, $\tau_D=\tau_U$), i.e., assuming
\begin{eqnarray}
\label{pdotdefs2}
\tau&=&\frac{\hbar/2}{\Boltz}\frac{\Delta_M}{\Delta_H}=
\frac{\hbar/2}{\Boltz}\sqrt{\frac{\scov{S}{S}}{\scov{H}{H}}-\frac{1}{\theta^2}}
\ ,\\ \scov{S}{S}&=&\Boltz^2\sum_n p_n(\ln p_n)^2-\mean{S}^2 \ .
\end{eqnarray}

The system of ordinary differential
 Equations (\ref{pdot}) is highly nonlinear, especially when $\tau$
  is assumed according to
 (\ref{pdotdefs2}), nevertheless it is sufficiently well
 behaved to allow simple integration by means of a standard
 Runge--Kutta numerical scheme. Of course, we check that at all
 times $-\infty<t<\infty$ each $p_n$ remains nonnegative, $\sum_n
 p_n$ remains equal to unity, $\sum_n p_ne_n$ remains constant at
 the value $2u/5$ fixed by the selected initial state, and the
 rate of change of $\mean{S}$ is always nonnegative.
\begin{figure}[h]
	\centering
		\subfloat[ $\tau$ constant]{{\includegraphics[width=0.38\textwidth]{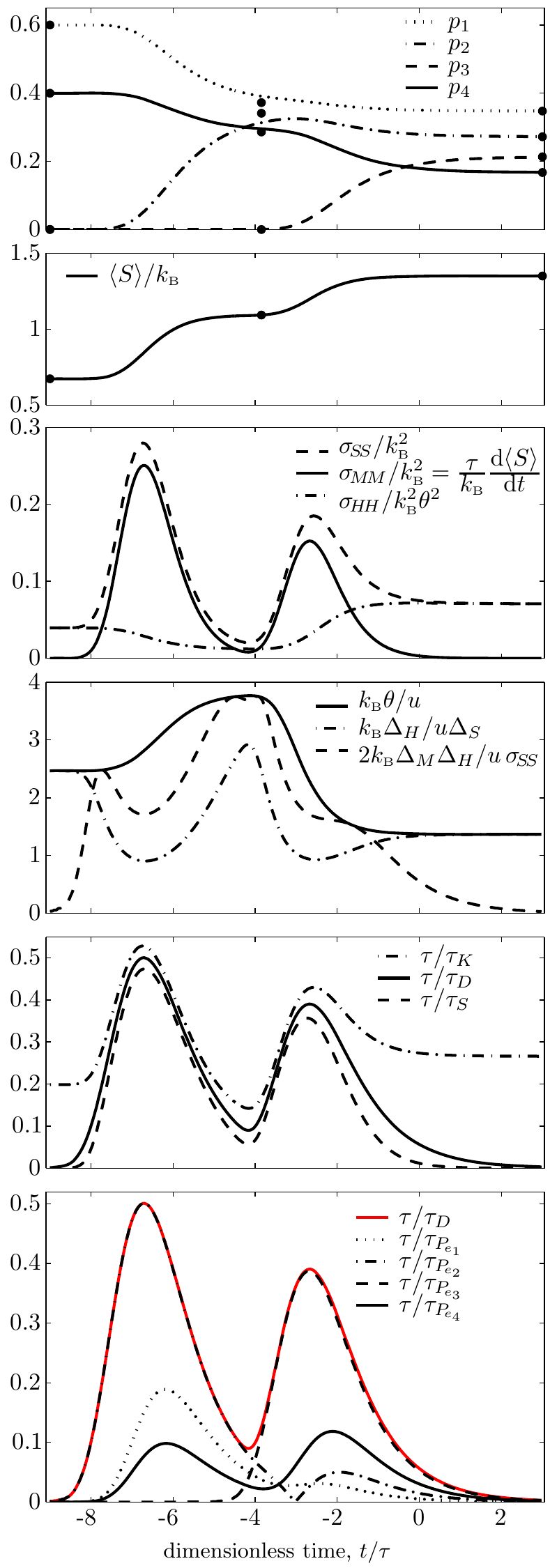}}}
		\qquad
		\subfloat[$\tau$ variable via Equation\,(\ref{pdotdefs2})]{{\includegraphics[width=0.38\textwidth]{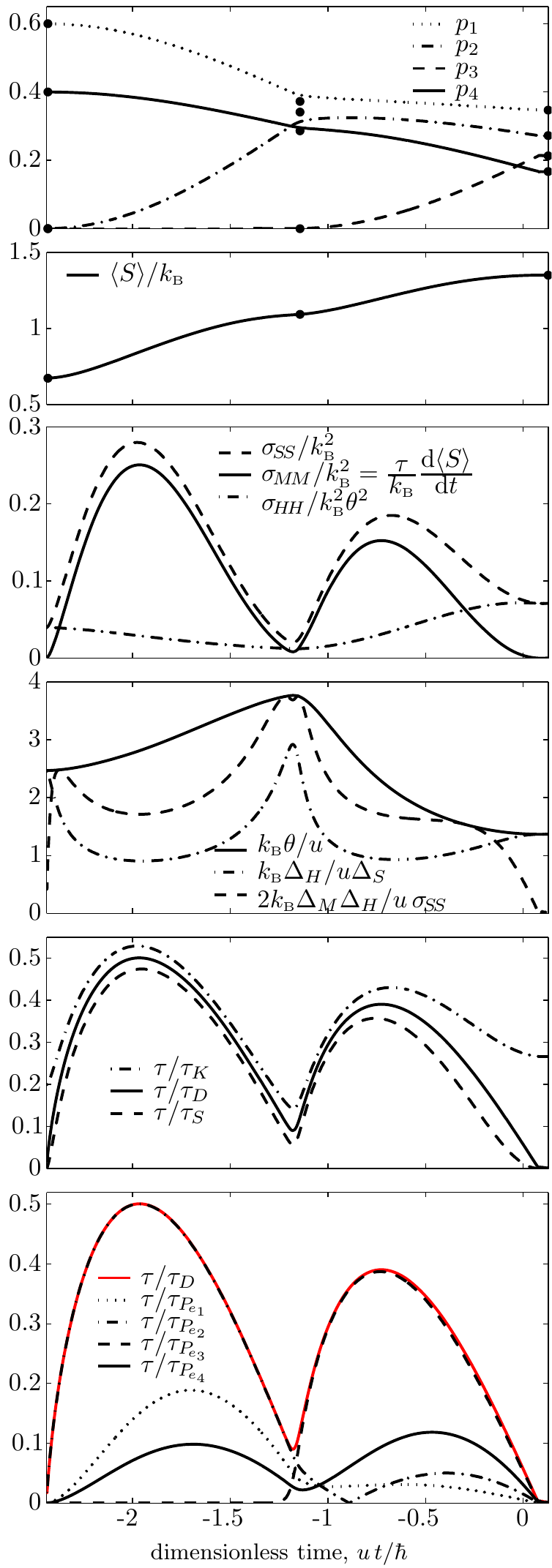}}}
		\caption{\label{Figure1}(color online) Time-dependent relaxation  results obtained by integrating the steepest-entropy-
		ascent master Equation (\ref{pdot}) for the four-level qudit with equally spaced energy levels, for two different choices of $\tau$: (\textbf{a}) $\tau=\rm const$ and (\textbf{b}) $\tau$ state-dependent according to Equation\ (\ref{pdotdefs2}) and time non-dimensionalized by $\hbar/u$ where $u$ is the energy difference between the highest and lowest energy levels of the system. First row subfigures: Time evolution of the four occupation probabilities $p_n$. Second row: dimensionless entropy $\mean{S}/\Boltz$. Third row:  rate of entropy
			change (proportional to $\scov{M}{M}$) compared with $\scov{S}{S}$
			and $\scov{H}{H}/\theta^2$, to illustrate relation (\ref{covSM}). Fourth row: generalized `nonequilibrium
			temperature' $\theta$ (nondimensionalized by $u/\Boltz$) compared  with $\Delta_H/\Delta_S$ and $2\Delta_M\Delta_H/\scov{S}{S}$ (also nondimensionalized)
			to illustrate relations (\ref{thetabound}) and
			(\ref{thetabound2}). Fifth row: characteristic time of purely dissipative evolution $\tau_D$ (here proportional to the inverse of the square
			root of the rate of entropy generation, shown in the third row subplots) compared
			with $\tau_S$ and $\tau_K$ to illustrate relations~\mbox{(\ref{tauSD}) and
			(\ref{deftauG})}. Sixth row:  characteristic times of the four occupation probabilities $\tau_\Pen$ compared with  $\tau_D$ to illustrate relation
			(\ref{teuPen}).}
\end{figure}
\unskip
\begin{figure}[h]
	\centering
	\subfloat[$\tau$  constant]{{
				\includegraphics[width=0.45\textwidth]{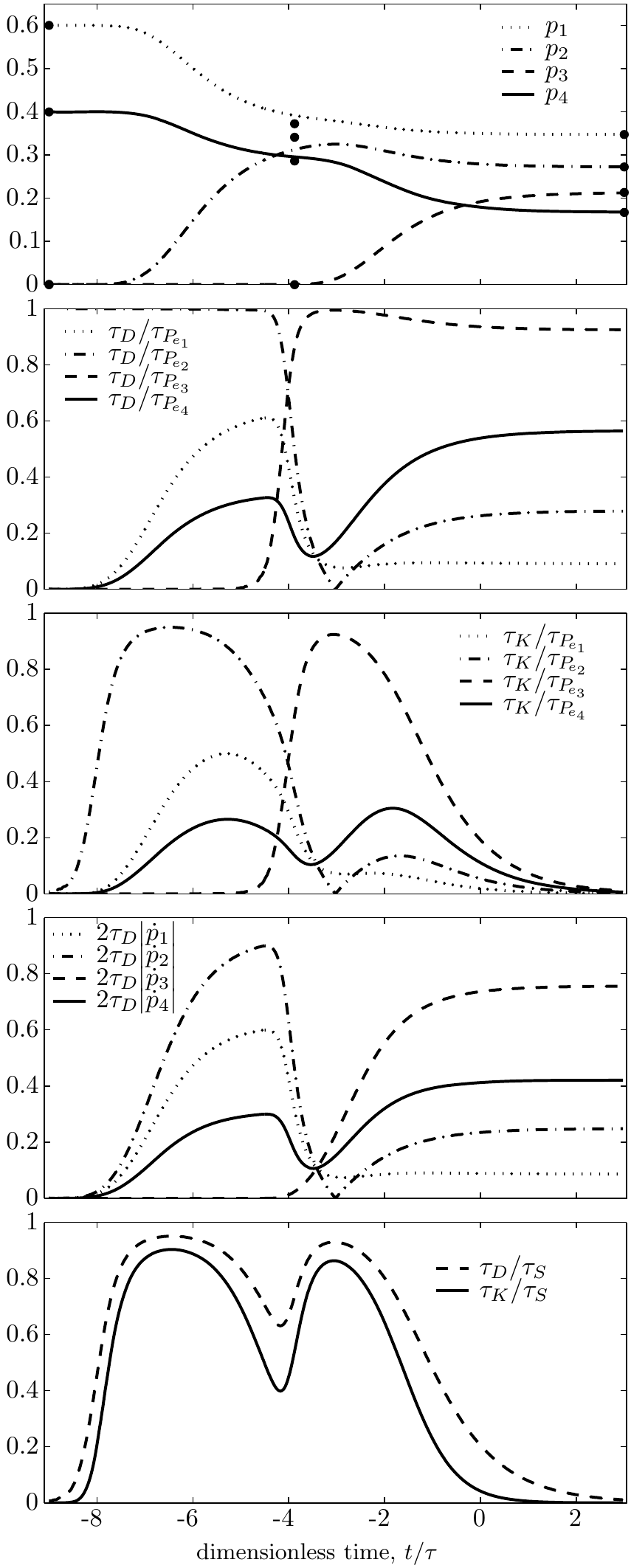}}}
		\qquad
		\subfloat[$\tau$ state-dependent via Equation\ (\ref{pdotdefs2})]{{\includegraphics[width=0.45\textwidth]{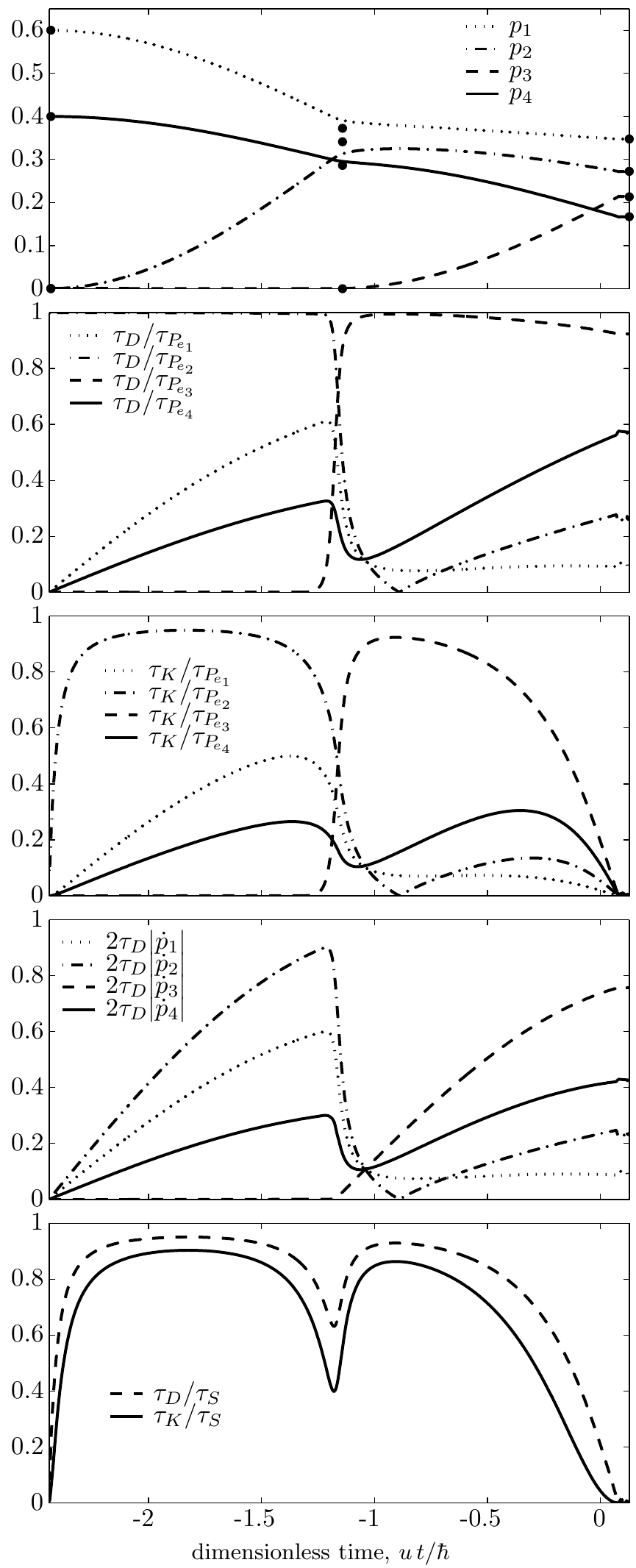}}}
		\caption{\label{Figure2}Time evolution of various other ratios of characteristic times for the same cases of Figure~\ref{Figure1}. First row subfigures: Time evolution of occupation probabilities $p_n$ (same as first row of Figure~\ref{Figure1}, repeated here for ease of comparison). 
			Second row: ratios $\tau_D/\tau_\Pen$ for each of the four occupation probabilities  to illustrate again relation  (\ref{teuPen}). Third row: $\tau_K/\tau_\Pen$  to illustrate relation  (\ref{genunc12}). Fourth row: $2\tau_D\,|\dot p_n|$  to illustrate relation (\ref{dpndt}). Fifth row: $\tau_D/\tau_S$ and $\tau_K/\tau_S$ to illustrate relations (\ref{tauSD}) and  (\ref{genunc5}).}
\end{figure}

  In  each Figure, the top subfigure shows for ease of comparison the
  plots of the four nonzero occupation probabilities as functions of dimensionless
  time:  $t/\tau$, in Figures \ref{Figure1}a and \ref{Figure2}a; $u\, t/\hbar$, in \mbox{Figures \ref{Figure1}b
  and \ref{Figure2}b}.
  The dots on the right represent the maximal
entropy distribution, $p_n(+\infty)=p_n^{\rm e}$; the dots
at the left represent the lowest-entropy or `primordial'
distribution, $p_n(-\infty)=p_{\rm nd}^{\rm ls}$, which for the
particular trajectory selected here, corresponds to a
nondissipative state $\rho_{\rm nd}^{\rm ls}$ that has only two
occupied energy levels, $e_1$ and $e_4$, with probabilities
$p_{{\rm nd}1}^{\rm ls}=0.6$ and $p_{{\rm nd}4}^{\rm ls}=0.4$ (and
temperature $T_{\rm nd}^{\rm ls}=2.466\,u/\Boltz $); in fact
the four-level system has no lower entropy states $\rho$ that commute with
$H$, have energy $2u/5$, and have zero occupation probabilities
 \cite{PRE06}. The dots in the middle represent
the nondissipative state $\rho_{\rm nd}^{\rm ft}$ which appears as
the false target state during the first part of the trajectory,
plotted at the instant in time when the entropy of the
time-varying trajectory is equal to the entropy of this
 distribution.

 It is interesting to observe from Figure \ref{Figure1} (bottom subfigures)
  that during the early
 part of the trajectory, $\tau_D$ almost exactly coincides with
 $\tau_\Petwo$ while in the late part it almost exactly coincides with
 $\tau_\Pethree$, and the switch occurs when the trajectory slows
 down in the neighborhood of the `false target' nondissipative
 state.

 In Figure \ref{Figure1}, the second subfigures show the time dependence of
 the dimensionless entropy $\mean{S}/\Boltz$; the third subfigures show its rate of
 change (proportional to $\scov{M}{M}$) and compares it with $\scov{S}{S}$
 and $\scov{H}{H}/\theta^2$, to illustrate relation (\ref{covSM}); the
 fourth show the time dependence of our generalized `nonequilibrium
 temperature' $\theta$ (properly nondimensionalized) and compares
 it with $\Delta_H/\Delta_S$ and $2\Delta_M\Delta_H/\scov{S}{S}$
 to illustrate relations (\ref{thetabound}) and
 (\ref{thetabound2}); the fifth subfigures show the time
 dependence of $1/\tau_D$ (which here is proportional to the square
 root of the rate of entropy generation, third subfigures) and compares
 it with $1/\tau_S$ and $1/\tau_K$ to illustrate relations
  (\ref{tauSD}) and (\ref{deftauG}); the sixth subfigures show
 $1/\tau_\Pen$ for each of the four occupation probabilities and
 compares them with  $1/\tau_D$ to illustrate relation
 (\ref{teuPen}), which for this particular trajectory has the
 feature we just discussed.

 In Figure \ref{Figure2}, the second subfigures illustrate again relation
 (\ref{teuPen}) for each of the four observables
 $p_n=\mean{\Pen}$; the third subfigures illustrate the time--entropy uncertainty relation
 (\ref{genunc12}) for the same observables; the fourth illustrate
 inequality (\ref{dpndt}); the fifth illustrate relations (\ref{tauSD}) and
 (\ref{genunc5}).

 By comparing subfigures (a) and (b) in both Figures \ref{Figure1} and \ref{Figure2}, it is noted 
 that most qualitative features remain the same when $\tau$ is changed from constant to the state-dependent functional defined by Equation\ (\ref{pdotdefs2}), except for the
 almost
 singular behavior near the false target partially canonical nondissipative state,
 where  $\Delta_M$ approaches zero and so does the dissipative time  $\tau$
 [Equation\ (\ref{pdotdefs2})]. The approach
 to final equilibrium in this case is not exponential in time as for
 $\tau =$ const. This puzzling behavior suggests that
  assumption (\ref{pdotdefs2}) may hardly be
 physically sensible. However, as already noted after (\ref{dissTEA}), it represents an interesting
 extreme behavior, i.e., the minimum dissipative time functional $\tau $
  by which observables
 that commute with $H$, like the occupations $\Pen$,  never violate the usual time--energy
 uncertainty relations $\tau_\Pen\Delta_H\ge \hbar/2$, even though
 their time dependence is not determined here by unitary dynamics but
 by purely dissipative dynamics.   These usual time--energy
 uncertainty relations, $\tau_\Pen\ge\tau_U$, are illustrated by the second row subfigures of Figure\ \ref{Figure2}, because
  in this case $\tau_U=\tau_D$.

\section{\label{Conclusions}Conclusions}

The  Mandelstam--Tamm--Messiah time--energy uncertainty relation
    $\tau_{F}\Delta_H\ge \hbar/2$ provides a general lower bound
    to the characteristic times of change of all observables of a
    quantum system that can be expressed as linear functionals of
    the density operator $\rho$. This has been used to obtain
    estimates of rates of change and lifetimes of unstable states,
    without  explicitly solving the time dependent evolution
    equation of the system. It may also be used as a general
    consistency check in measurements of time dependent phenomena.
    In this respect, the exact relation and inequalities
    (\ref{exactTE}) [that we derive for standard unitary dynamics based
    on the generalized Schr\"odinger inequality
    (\ref{Sinequality})] provide, for unitary evolution, a more general and sharper chain
    of consistency checks than the usual time--energy uncertainty relation.

    The growing interest during the last three or four decades in
    quantum dynamical models of systems undergoing irreversible
    processes has been motivated by impressive technological
    advances in the manipulation of smaller and smaller systems,
    from the micrometer scale to the nanometer scale, and down to
    the single atom scale. The laws of thermodynamics, that fifty
    years ago were invariably understood as pertaining only to
    macroscopic phenomena, have gradually earned more attention and
    a central role in studies of mesoscopic phenomena first, and of
    microscopic and quantum phenomena more recently. In this paper we do not address
     the controversial issues currently under discussion
    about interpretational matters, nor do we attempt a reconstruction and
    review of the different views,  detailed models and  pioneering contributions
    that propelled during the past two decades this fascinating advance of thermodynamics
    towards the realm of few particle and single particle systems.

    Motivated by this context and background, we derive various
    extensions of the usual time--energy uncertainty relations that
     may become useful in phenomenological studies of dissipative phenomena.
     We do so by focusing
 on a special but broad class of model evolution
    equations, that  has been designed for the description of dissipative
    quantum phenomena and for satisfying a set of strict compatibility conditions
    with  general thermodynamic principles. In this framework, we derive various forms of considerably precise
    time--energy and time--entropy
    uncertainty relations, and other interesting general inequalities, that
    should turn out to be  useful at least as additional
    consistency checks in measurements of nonequilibrium states and time-dependent dissipative phenomena. To illustrate the qualitative features and the sharpness of  the bounds provided by this set of inequalities,     
   we show
    and discuss a numerical example obtained by integration (forward and backward in time) of the nonlinear evolution
    equation in the specific form  introduced by this author for the description of
    steepest-entropy-ascent dynamics of an isolated system far
    from thermodynamic equilibrium.

  \vspace{6pt}
    

\section*{Acknowledgements}

The author is grateful to Victor Dodonov for an interesting discussion during his visit in Brescia on 11 July 2007 on the viewgraphs I had just presented at the Conference on ``Quantum Theory: Reconsideration of Foundations--4'' in 	Växjö, Sweden, 11--16 June 2007, based on the  early version of the present paper  I had uploaded in ArXiv in 2005 \cite{ArXiv05}, also presented at the 10th International Conference on ``Squeezed States and Uncertainty Relations,'' Bradford, UK, 2 April 2007 with the title  ``Time-Energy and Time-Entropy Uncertainty Relations in Steepest-Entropy-Ascent Dissipative Dynamics.'' For the sake of historical record, early versions of this paper were submitted to Physical Review Letters (LL10220, December 2005) and to Physical Review A (LL10220A, April 2006) but were rejected after five mixed peer reviews. 
    

\appendix

\section{\label{AppendixA}Reasons for Not Assuming a
Kossakowski--Lindblad form of the Master Equation}

With various motivations, fundamental or phenomenological, dissipative quantum dynamical models, i.e., evolution equations for the density operator
$\rho$ that do not conserve the functional $-\Tr(\rho\ln\rho)$,  are almost invariably based on the KLGS master equations. For example, in  theories of open systems in contact with a heat bath, or
subsystems of a composite system which as a whole evolves unitarily,
a variety of successful model evolution equations for the reduced
density operator of the system have the KLGS form \cite{Kraus71,Kossakowski72_1,Kossakowski72_2,Ingarden75,Lindblad76,Gorini76,Spohn76,Spohn78,Asorey09}
\begin{equation}
\label{Lindblad}
\ddt{\rho}=-\frac{i}{\hbar}[H,\rho]+\tfrac{1}{2}\sum_j\left(2
V_j^\dagger\rho  V_j -\{ V_j^\dagger V_j,\rho\}\right) \ ,
\end{equation}
where the $V_j$'s are operators on $\Hil$ (each term within the
summation, often written in the alternative form $[V_j,\rho
V_j^\dagger]+ [V_j\rho, V_j^\dagger]$, is obviously traceless).
Evolution equations of this form are linear in the density
operator $\rho$ and preserve its hermiticity, nonnegativity and
trace.  

For example, in a number of successful models of dissipative
quantum dynamics of open subsystems,  operators $V_j$ are in
general interpreted as creation and annihilation, or transition
operators. For example, by choosing $V_j=c_{rs}|r\rangle\langle
s|$,  where $ c_{rs} $ are complex scalars and $ | s\rangle $
eigenvectors of the Hamiltonian operator $H$, and defining the
transition probabilities $w_{r s}= c_{rs} c_{rs}^*  $, Equation
(\ref{Lindblad}) becomes
\begin{equation}
\label{Pauli1} \ddt{\rho} = - \, \fr{i}{\hbar} [ H , \, \rho ] +
\sum_{r s} w_{r s} \left( | s \rangle \langle\rho \rangle\langle s
| - \, \fr{1}{2} \{ | s \rangle\langle s | \, , \rho \} \right) \,
,
\end{equation}
or, equivalently, for the $nm$-th matrix element of $\rho$ in
the $H$ representation,
\begin{equation}
\label{Pauli2} \ddt{\rho_{nm}} =
 -\fr{i}{\hbar} \rho_{nm} ( E_n - E_m )  
 + \delta_{nm}
 \sum_r w_{n r} \rho_{r
r} - \rho_{nm} \fr{1}{2} \sum_r ( w_{r n} + w_{r m} ) \ ,
\end{equation}
which, for the occupation probabilities $p_n=\rho_{nn}$, is the
Pauli master equation
\begin{equation}\label{Pauli3} \ddt{ p_n} = \sum_r w_{n r} p_r -
p_n \sum_r w_{r n} \  .
\end{equation}

Equation
(\ref{Lindblad}) has  also the
intriguing feature of generating a completely positive dynamical map. However,  Reference \cite{Simmons81_2}
argues quite clearly that the requirement of
complete positivity of the reduced dynamics is too restrictive, as
it is physically unnecessary to assure preservation of positivity
of the density operator of the composite of any two
noninteracting, uncorrelated systems.

Our objective here, instead, is to consider a  class of model evolution equations
applicable not only to  open systems but also to
 closed isolated systems, capable of describing, simultaneously with the usual
Hamiltonian unitary evolution, the natural tendency of any initial
nonequilibrium state to relax towards canonical or
partially-canonical thermodynamic equilibrium, i.e., capable of
describing the irreversible tendency to evolve towards the highest
entropy state compatible with the instantaneous mean values of the
energy, the other constants of the motion, and possibly other
constraints. To avoid the severe restrictions imposed by the
linearity of the evolution equation, we open our attention to
nonlinearity in the density operator $\rho$ \cite{Simmons81}.
Therefore, it may at first appear natural to maintain the
Kossakowski--Lindblad form (\ref{Lindblad}) and simply assume 
operators $V_j$ that are functions of $\rho$. This is true only in part
for the evolution Equation (\ref{rhodotHM}) that we assume.
Indeed, our hermitian operator $\Delta M(\rho)/\Boltz\tau $
can always be written as $-\sum_j V_j^\dagger(\rho) V_j(\rho)$ and
therefore our anticommutator term may be viewed as a
generalization of the corresponding term in (\ref{Lindblad}).

However, in our Equations (\ref{rhodot}) and  (\ref{rhodotHM}) we
suppress the term corresponding to  $\sum_j V_j^\dagger\rho V_j$
in (\ref{Lindblad}). The reason for this suppression is the
following. Due to the terms $V_j^\dagger\rho V_j$, whenever the
state operator $\rho$ is singular, i.e., it has one or more zero
eigenvalues, Equation (\ref{Lindblad})  implies that these zero
eigenvalues may change at a finite rate. This can be seen clearly
from (\ref{Pauli3}) by which ${\rm d}p_n/{\rm d}t$ is  finite
whenever there is a nonzero transition probability $w_{nr}$ from
some other populated level ($p_r\ne 0$), regardless of whether
$p_n$ is zero or not. When this occurs, for one instant in time
the rate of entropy change is infinite, as seen clearly from the
expression of the rate of entropy change implied by
(\ref{Lindblad}),
\begin{equation}
\label{PauliS1} \ddt{\mean{S}} = \Boltz \sum_j
\Tr ( V_j^{\dagger}
 V_j \rho \ln
\rho - V_j^{\dagger} \rho V_j \ln \rho )= \Boltz
\sum_{j r n}
 ( V_j )^*_{n r}  ( V_j )_{n r} ( \rho_r - \rho_n ) \ln \rho_r \ ,
\end{equation}
where $\rho_r$ denotes the $r$-th eigenvalue of $\rho$ and $( V_j
)_{n r}$ the matrix elements of $V_j$ in the $\rho$
representation.

We may argue that an infinite rate of entropy change  can  be
tolerated, because it would last only for one instant in time. But
the fact that zero eigenvalues of $\rho$ in general could not
survive, i.e., would not remain zero (or close to zero) for longer
than one instant in time, is an unphysical feature, at least
because it is in contrast with a wealth of successful models of
physical systems in which great simplification is achieved  by
limiting our attention to a restricted subset of relevant
eigenstates (forming a subspace of $\Hil$ that we call the
effective Hilbert space of the system \cite{MPLA1}). Such common
practice $N$-level models yield extremely good results, that being
reproducible, ought to be relatively robust with respect to
including in the model other less relevant eigenstates. In fact,
such added
 eigenstates, when initially unpopulated, are irrelevant if
 they remain unpopulated
(or very little populated) for long times, so that neglecting
their existence introduces very little error. The terms
$V_j^\dagger\rho V_j$, instead, would rapidly populate such
irrelevant unpopulated eigenstates and void the validity of our so
successful simple $N$-level models, unless we deliberately  overlook this
instability problem by highly ad-hoc assumption, e.g., by forcing
the $V_j$'s to be such that $( V_j )_{n r}=0$ whenever either
$\rho_n=0$ or $\rho_r=0$, in which case, however, we can no longer
claim true linearity with respect to $\rho$.

To avoid the unphysical implications of this seldom recognized
\cite{MPLA1,PRE06} problem of linear evolution equations of form
(\ref{Lindblad}), we consider in this paper only equations of form
(\ref{rhodotHM}). We do not exclude that it may be interesting to
investigate also the behavior of equations that include  nonlinear
terms of the form $V_j^\dagger(\rho)\, \rho\, V_j(\rho)$. However,
at least when the system is strictly isolated, the
operator-functions $V_j(\rho)$ should be such that $( V_j(\rho)
)_{n r}=0$ whenever either $\rho_n=0$ or $\rho_r=0$.

Another important general physical reason why we exclude terms that
generate nonzero rates of change of zero eigenvalues of $\rho$, is
that if such terms are construed so as to conserve positivity in
forward time, in general they cannot maintain positivity in backward
time. The view implicitly assumed when Equation (\ref{Lindblad}) is
adopted, is that the model is ``mathematically irreversible'' (a
distinguishing feature if not a starting point of the theory of
completely positive linear dynamical semigroups on which it is
based), in the sense that neither uniqueness of solutions in forward
time nor existence in backward time are required (and granted). Such
mathematical irreversibility of the initial value problem, is often
accepted, presented and justified as a natural counterpart of
physical irreversibility. However, it is more related to the
principle of causality than to physical irreversibility. The
strongest form of the non-relativistic principle of causality---a
keystone of traditional physical thought---requires that future
states of a system should unfold deterministically from initial
states along smooth unique trajectories in state domain defined for
all times (future as well as past). Accepting mathematical
irreversibility of the model dynamics implies giving up such
causality requirement. The point is that such requirement
 is not strictly necessary  to describe physical irreversibility,
 at least not if we are willing  to give up linearity instead. The
proof of this statement is our Equation (\ref{rhodotHM}) which,
together with the additional assumptions made in Section
\ref{Example} to describe relaxation within an isolated system, is
mathematically reversible, in the sense that it features existence
and uniqueness of well-defined solutions both in forward and
backward time, and yet it does describe physically irreversible
time evolutions, in the sense that the physical property described
by the entropy functional $-\Boltz\Tr(\rho\ln\rho)$ is a strictly
increasing function of time for all states except  the very
restricted subset defined by Equation~(\ref{rhonondiss}), where it is
time invariant.

\section{\label{AppendixB}How Did Locally Steepest Entropy Ascent Come About?}

The KLGS master equation emerges from a bottom-up phenomenological approach, whereby one considers a weakly interacting  system$+$bath isolated composite evolving under the phenomenological assumption that the large number of degrees of freedom of the bath dilutes and destroys the  correlations that build up under the standard  unitary evolution due to the interaction term in the Hamiltonian, so that for the purpose of computing the evolution of the reduced density operator $\rho_S$ of the system, the overall state can be assumed to evolve through uncorrelated states, i.e., the system$+$bath density operator can at all times be written as $\rho=\rho_S\otimes\rho_B$. So, we may say that the derivation is ``bottom-up'' because it starts from the fundamental unitary  evolution of the composite system, but it is also ``phenomenological'' because the assumption of loss of correlations is an approximation that depends on the bath details and requires neglecting some terms during the partial tracing over the bath~subspace.

By contrast, the locally steepest-entropy-ascent master equation was originally constructed (not derived) from a ``top-down'' approach, meant to see what modifications of standard quantum mechanics would be required if one wants to embed the second law of thermodynamics directly into the fundamental law of description.  This heretic, but certainly intriguing and thought-provoking theoretical exercise, belongs to the early history of quantum thermodynamics and should not be forgotten.   In the 70's, the need for a quantum thermodynamics had been addressed boldly and explicitly only by Hatsopoulos and Gyftopoulos \cite{HG3,HG1,HG2,HG4} and---with a very different approach that we do not review here---by Prigogine and the Brussels school of thermodynamics \cite{Prigogine77,Prigogine78,Prigogine79}. As 
an additional note pertaining to the history of thermodynamics, the course 2.47~J/22.58~J,  listed in the MIT Bulletin of the academic year 1970-71 and taught jointly by George N. Hatsopoulos and Elias P. Gyftopoulos in the Spring of 1971, is the first official course entitled ``Quantum Thermodynamics'' that we are aware of.

An emphatic way to explain the (philosophical?) intuition of these pioneers of quantum thermodynamics, is the set of ``what if'' questions posed by the present author in a famous conference on the frontiers of nonequilibrium statistical physics held in Santa Fe in 1984 \cite{SantaFe84_2}:
``what if entropy, rather than a statistical, information
theoretic, macroscopic or phenomenological concept, were an intrinsic
property of matter in the same sense as energy is universally
understood to be an intrinsic property of matter? What if
irreversibility were an intrinsic feature of the fundamental
dynamical laws obeyed by all physical objects, macroscopic and
microscopic, complex and simple, large and small? What if the second
law of thermodynamics, in the hierarchy of physical laws, were at the
same level as the fundamental laws of mechanics, such as the great
conservation principles? Is it inevitable that the gap between
mechanics and thermodynamics be bridged by resorting to the usual
statistical, phenomenological, or information-theoretic reasoning,
and by hinging on the hardly definable distinction between
microscopic and macroscopic reality? Is it inevitable that
irreversibility be explained by designing ad hoc mechanisms of
coupling with some heat bath, reservoir or environment, and ad hoc
mechanisms of loss of correlation? What if, instead, mechanics and
thermodynamics were both special cases of a more general unified
fundamental physical theory valid for all systems, including a single
strictly isolated particle, such as a single isolated harmonic
oscillator or a single isolated two-level spin system?''

In References \cite{HG3,HG1,HG2,HG4} Hatsopoulos and Gyftopoulos showed that the only price  we have to pay \cite{Perimeter07,Perimeter09} to gain a possible positive answer to these questions is the  reinterpretation of the physical meaning of the density operator, abandoning the standard (statistical mechanics and information theoretic) interpretation whereby it represents
 the epistemic ignorance of which particular pure state the
system is `really' in. Instead, the density operator  acquires an `ontic' status and  represents the individual state
of the isolated and uncorrelated atom (or particle or indivisible entity of the system's model). Equivalently, in terms
of ensembles, the density operator represents the measurement
statistics from a {homogeneous ensemble}---homogeneous in
the sense defined by von Neumann and discussed in References
\cite{vonNeumann,ParkState,thesis,MPLA2}, i.e., such that no
subsensemble can be identified which gives rise to
{different} measurement~statistics.

As a result of such ansatz, the severe
restrictions imposed by linearity on the evolution equation become
unnecessary,
and we must open up our attention to evolution equations nonlinear in the density operator $\rho$. This is what prompted the search for a   generalization of the time-dependent Schr\"odinger
equation for pure density operators, to a broader fundamental kinematics in which not only every pure density
operator, but also every non-pure density operator represents a
real ontological object, the `true' state of the system, which can be mixed even if the system is isolated and uncorrelated (no entanglement) from the rest of the universe. Among the desiderata \cite{MPLA1} that drove the search for an extension, the most important was that the second law should emerge as a theorem of the equation of motion, which we considered the strongest way to enforce strong compatibility of a dynamical model (or law of motion or resource theory\dots) with thermodynamics.

 For such purpose, the  Hatsopoulos--Keenan statement of the second law  is particularly suited, because it is directly linked with stability features of the equilibrium states of the dynamics.   This somewhat still overlooked statement of the second law asserts \cite[p.~62]{Book} that for any well-defined (i.e., separable and uncorrelated) system, among the set of states that share the same values of the parameters of the Hamiltonian and the (mean) values of the  energy and the amounts of constituents, there exists one and only one (conditionally~\cite{Lyapunov,Hiai}) stable equilibrium state, which turns out to be the one with the maximal entropy, often called the Gibbs state in the recent QT literature.

The  Hatsopoulos--Keenan statement of the second law not only can be proved to entail the better known statements  (Kelvin--Planck \cite[p.~64]{Book}, Clausius \cite[p.~134]{Book}, Caratheodory \cite[p.~121]{Book}), but---quite importantly for the current developments of  quantum thermodynamics---it supports a rigorous operational definition of entropy as a general property of any uncorrelated (and unentangled) state of any well-separated system, valid not only for the stable equilibrium states of macroscopic systems but also for their nonequilibrium states (see Reference \cite{AAPP19} and references therein) and providing a possible  basis  for its extension to  systems  with only few particles and quantum systems. Its~extendability to correlated states of interacting or non-interacting systems is instead still the subject of intense debate, because the correlation entropy (often called mutual information), like the mean energy of interaction between the subsystems, is a well defined feature of the overall state of a composite system, but  there is no unique way nor fundamental reason to allocate it among the subsystems and assign it to their local (reduced, marginal) states.

Using a similar operational definition, Hatsopoulos and Gyftopoulos in References \cite{HG3,HG1,HG2,HG4} showed that  the von Neumann  entropy functional $-\Boltz\Tr(\rho\ln\rho)$ fulfills the definition of entropy. 
It was for such pioneering QT framework that the present author designed  the nonlinear master equation \cite{Cimento1,Cimento2,Nature} and shortly thereafter \cite{Maryland86} proved it admits a steepest-entropy-ascent (SEA) variational formulation that embodies at the local microscopic level the principle of maximal entropy production~\cite{Sieniutycz87,Gheorghiu1,Dewar05,Martyushev06,PRE06,ROMP,Bregenz09}. 

Starting in the mid eighties, during times when quantum thermodynamics was considered interesting only by  a handful of pioneers, the present author has addressed both the  quantum-foundations and mathematical--physics  communities \cite{SantaFe84_1,SantaFe84_2,Maryland86,Paris87,Taormina92,Messina05,IJQT07,Vaxjo07,Perimeter07,MIT08H,Perimeter09,Bregenz09,Aachen18} and the engineering-thermodynamics and non-equilibrium-thermodynamics communities \cite{Anaheim86,Boston87,Roma87,Leiden06,StEtienne07,MIT08,SanDiego13}
to raise awareness about the requirement that  fundamental and  phenomenological models of dissipative and irreversible processes must incorporate, i.e., must not violate, general thermodynamic principles. 

More recently, the SEA principle has been shown to encompass all the major levels of  description of nonequilibrium dynamics and irreversible processes \cite{PRE06}, including the general modeling structure known as metriplectic dynamics \cite{Kaufman84,Morrison84,Grmela84} or GENERIC  
(see  \cite{PRE15} and references therein), which even more recently has been shown to bear deep connections also with the mathematical theories of gradient flows \cite{Jordan98,Otto01,Mielke11} and large fluctuations \cite{Peletier14,Montefusco18}.

In other words, the locally steepest-entropy-ascent model of far-non-equilibrium dissipative evolution in QT can be considered the most general precursor of all more recent and successful  theories of nonequilibrium dynamical systems.

In several instances and different fields of application the LSEA approach has shown the ability to provide new nontrivial modeling capabilities not only in the realm of QT and related phenomenological resource theories (see, e.g., References \cite{Cano15,Tabakin17,Militello18}) but also in materials science \cite{Kusaba17,Yamada18,Yamada19,Yamada19PRE} and transport theory~\cite{Li18}.  

Also the unified theory presented in References \cite{HG3,HG1,HG2,HG4}, if one puts aside the epistemic interpretation and considers it as an effective resource theory, represents in our view the pioneering precursor of many quantum thermodynamics results that have been re-derived in recent years (free energy versus available energy, energy versus entropy diagram, work element, adiabatic availability, etc).


\end{document}